\documentclass[aps,prd,onecolumn,nofootinbib,superscriptaddress,11pt]{revtex4-2}
\usepackage[utf8]{inputenc}
\usepackage{amsmath,amssymb,graphicx,bm,slashed}
\usepackage[colorlinks=true,linkcolor=blue,citecolor=blue,urlcolor=blue]{hyperref}

\allowdisplaybreaks
\newcommand{\MPl}{M_{\rm Pl}}
\newcommand{\cFa}{c^{a}_{F^4}}
\newcommand{\cFb}{c^{b}_{F^4}}
\newcommand{\cD}{c_{D^4\phi^4}}
\newcommand{\cDFa}{c^{a}_{D^2\phi^2F^2}}
\newcommand{\cDFb}{c^{b}_{D^2\phi^2F^2}}
\newcommand{\cPF}{c_{\Psi^2F^2}}
\newcommand{\cDP}{c_{D^2\phi^2\Psi^2}}
\newcommand{\gFa}{\gamma^{a}_{F^4}}
\newcommand{\gFb}{\gamma^{b}_{F^4}}
\newcommand{\ang}[1]{\langle #1\rangle}
\newcommand{\sq}[1]{[#1]}

\begin{document}

\begin{flushright}
		IFT-UAM/CSIC-26-164\\
\end{flushright}

\title{Positivity Bounds, Negative Running, and the Weak Gravity Conjecture}

\author{Jaime Fernandez}
\email{jaime.fernandez@csic.es}
\affiliation{Instituto de F\'{\i}sica Te\'orica UAM/CSIC, Madrid 28049, Spain}

\begin{abstract}
    We study causality and unitarity constraints at one loop on a low-energy effective theory containing a photon, a massless scalar coupled to it through dilaton-like interactions, and a light neutral Dirac fermion with a Yukawa coupling to the scalar. In this setup, non-gravitational loop effects induced by the Yukawa interaction can drive the coefficient of the four-photon $(F_{\mu\nu}F^{\mu\nu})^2$ operator negative at low energies, apparently violating the tree-level positivity bound. Nevertheless, we find that these negative values are consistent with one-loop constraints from causality and unitarity, thanks to the low-energy contribution from the massless scalar to the relevant dispersion relation. We also examine the implications of this negative running for the Weak Gravity Conjecture in the magnetic limit of Einstein--Maxwell--dilaton black holes, since the RG-evolved Wilson coefficients can a priori shift the extremality condition towards the subextremal region. We find, however, that this shift is obstructed by the fermion mass induced by the same Yukawa coupling, which prevents the fermion from contributing to the running of the relevant operators down to the low-energy scales required to make the black holes subextremal. Finally, we discuss the relevance of negative running to loop-level positivity bounds in the SMEFT and to light-by-light scattering in the presence of a light scalar.
\end{abstract}

\maketitle
\tableofcontents

\section{Introduction}
\label{sec:intro}

Causality, unitarity and analyticity of the $S$-matrix constrain the Wilson coefficients (WCs) of low-energy effective field theories (EFTs). At tree level, these \emph{positivity bounds} state, for instance, that the coefficients of the leading four-derivative self-interactions of a shift-symmetric scalar or of a photon must be positive~\cite{Adams1}. Positivity bounds have found applications ranging from collider phenomenology~\cite{Bellazzini1,Bellazzini2,Bi1,Yamashita1,Li1,Liu1,Fernandez1} to the swampland programme, where they have been used to argue for the Weak Gravity Conjecture (WGC)~\cite{ArkaniHamed1} in Einstein--Maxwell theory~\cite{Adams1,Kats1} and its extensions~\cite{Cheung1,Bellazzini3,Goon1,Hamada1,Bellazzini4,Charles1,ArkaniHamed2,Noumi1,McPeak1,Tokareva1,DiRusso1}.\footnote{For a comprehensive review of the WGC and the evidence supporting it, see Ref.~\cite{Harlow1}.}

Here we focus on positivity bounds beyond tree level. At loop level, the low-energy part of the dispersive integrals receives corrections that can be computed within the EFT itself~\cite{Bellazzini5}, while the ultraviolet (UV) side remains positive. Loop corrections to dispersive bounds have two relevant consequences. First, WCs acquire a scale dependence, and the bounds apply to combinations evaluated at a finite energy $E$ below the cutoff $\Lambda$~\cite{Bellazzini5,Chala1,Li2,Chang1}. Second, light states can produce infrared (IR) terms that relax the bounds, as happens for graviton exchange~\cite{Henriksson1,CaronHuot1,Chang1}, for instance. A particularly compelling scenario is that in which WCs become increasingly negative at lower energies due to their renormalization-group (RG) evolution, which apparently violates the tree-level bounds at sufficiently low scales. This raises the question of which interactions can produce negative running (i.e., positive beta functions). In massless theories, minimal interactions have been shown to contribute positively to the running of the WCs of the least irrelevant operators entering elastic amplitudes of four photons and four shift-symmetric scalars, and the same is true for leading-order non-minimal interactions in isolation~\cite{Baratella1,Fernandez2}; within this setup, negative running can arise only from interference between minimal and leading EFT interactions~\cite{Fernandez2}. A straightforward mechanism for negative running arises from terms in the beta function that are linear in both an EFT coefficient and a renormalizable coupling, since such contributions can have either sign~\cite{Chala1,Chala2}.

In this work, we study a simple system in which this possibility is realized: a photon, a massless scalar $\phi$ coupled to it through $e^{-2\kappa\phi}F_{\mu\nu}F^{\mu\nu}$, and a light neutral Dirac fermion $\Psi$ with a Yukawa coupling $y\,\phi\bar\Psi\Psi$. This is the decoupling limit of Einstein--Maxwell--dilaton (EMd) theory~\cite{Gibbons1,Garfinkle1,Cheng1} extended by a fermion. In this theory, we find that the interference between the Yukawa interaction and a fermionic two-photon EFT operator induces a sign-indefinite contribution to the running of the four-photon $(F_{\mu\nu}F^{\mu\nu})^2$ coefficient. Furthermore, we find that allowing this coefficient to run to arbitrarily negative values does not necessarily violate causality and unitarity constraints, thanks to fermion-loop corrections to the scalar exchange, which enter the relevant dispersion relation.

This result can be of relevance for the applications that we described before. Here we mainly focus on its consequences for the WGC. Negative running has previously been studied in the context of extremal black-hole (BH) decay~\cite{Charles1,ArkaniHamed2}. In those studies, however, negative running arose from gravitational interactions. Therefore, to reverse the sign of a WC that is positive near the matching scale, one must evolve the coefficient down to scales that are exponentially suppressed by the ratio of the cutoff of the theory to $\MPl$. Consequently, BHs large enough to probe negative WCs are also sensitive to the cosmological constant, unless the latter is extremely fine-tuned~\cite{ArkaniHamed2}. In contrast, the negative running found in our theory does not arise from gravitational interactions and therefore avoids this particular fine-tuning problem. Substituting the vacuum RG-evolved WCs into the corrections to the extremality condition therefore leads to an apparent tension with the WGC. However, in our case, a more careful analysis of the BH background shows that the induced fermion mass obstructs the massless-fermion RG extrapolation. Remarkably, the very interaction responsible for the negative running also undermines the RG-based argument that this running could obstruct the decay of extremal black holes.

Beyond the WGC, we briefly discuss consequences of the negative running for loop-level positivity in the Standard Model EFT (SMEFT), where dimension-eight coefficients can likewise run with either sign, and for light-by-light scattering in the presence of light scalars. In the latter scenario, we consider whether a scalar like the one in our theory could remain undetected even if a negative four-photon coefficient were measured.

This paper is organized as follows. In Sec.~\ref{sec:eft} we present the EFT, the assumed coupling hierarchies, and our power-counting assumptions. Sec.~\ref{sec:positivity} contains the one-loop positivity bounds for the relevant dimension-eight operators and discusses the possibility of negative running. In Sec.~\ref{sec:bh} we analyze the implications for extremal black holes in EMd theory, including the apparent tension with the WGC and the effects of the background on the underlying argument. Sec.~\ref{sec:avenues} discusses other scenarios in which negative running compatible with positivity constraints may be relevant. We conclude with Sec.~\ref{sec:conclusions}.

\section{The effective theory}
\label{sec:eft}

The particle content of the theory we use in this work consists of a massless real scalar $\phi$, a $U(1)$ photon $A_\mu$, and two massless Weyl fermions $\psi_\alpha$, $\tilde\chi^{\dot\alpha}$ combined into a Dirac fermion $\Psi\equiv(\psi_\alpha,\tilde\chi^{\dot\alpha})$, in $d=4$ spacetime dimensions. Neither the scalar nor the fermion is charged under the $U(1)$. We introduce a dilaton-like coupling between the photon and the scalar, together with a Yukawa coupling between the fermion and the scalar. The resulting minimal action is
\begin{equation}
    S_{\rm min}=\int d^4x\sqrt{-g}\left[-\frac12R+\frac12\partial_\mu\phi\partial^\mu\phi-\frac14e^{-2\kappa\phi}F_{\mu\nu}F^{\mu\nu}+i\bar\Psi\slashed{D}\Psi+y\,\phi\,(\bar\Psi\Psi)\right],
    \label{eq:Smin}
\end{equation}
where we use the mostly-minus metric and set $\MPl^2\equiv1/(8\pi G_N)=1$. Notice that for $y=0$, $\phi$ plays the role of the dilaton of EMd theory in the Einstein frame. We therefore treat $y\ll1$ as a perturbation that breaks the shift symmetry of $\phi$, and we include no potential for $\phi$ (see the discussion in Sec.~\ref{sec:scalarpotential}). For the positivity analysis we further assume $\Lambda\ll\MPl$ and neglect gravity, but we still keep $\kappa\Lambda\lesssim1$ non-negligible. In Planck units, this corresponds to $\kappa\gg1$.

Regarding EFT interactions, the higher-dimensional operators relevant for this work are
\begin{align}
    \Delta\mathcal{L}_{\rm EFT}\supset{}&\cFa\,e^{-6\kappa\phi}(F_{\mu\nu}F^{\mu\nu})^2+\cFb\,e^{-6\kappa\phi}(F_{\mu\nu}\tilde F^{\mu\nu})^2+\cD\,e^{-2\kappa\phi}(\partial_\mu\phi\partial^\mu\phi)^2\nonumber\\
    &+\cDFa\,e^{-4\kappa\phi}\partial_\mu\phi\partial^\mu\phi F_{\nu\rho}F^{\nu\rho}+\cDFb\,e^{-4\kappa\phi}\partial_\mu\phi\partial_\nu\phi F^{\mu\rho}F^{\nu}{}_{\rho}\nonumber\\
    &+\cPF\,e^{-3\kappa\phi}(\bar\Psi\Psi)F_{\mu\nu}F^{\mu\nu}+\cDP\,e^{-\kappa\phi}(\partial_\mu\phi\partial^\mu\phi)\bar\Psi\Psi\,,
    \label{eq:LEFT}
\end{align}
with $\tilde F_{\mu\nu}\equiv\frac12e^{-2\kappa\phi}\epsilon_{\mu\nu\rho\sigma}F^{\rho\sigma}$.\footnote{Since we regard gravity as negligible for the positivity analysis, we omit the four-derivative operators involving curvature tensors in Eq.~\eqref{eq:LEFT}, which are nevertheless considered in related works~\cite{Loges1,Bellazzini4}.} The dilaton exponentials are fixed by requiring the dilaton dependence to factorize in the string frame~\cite{Ibanez1}.\footnote{The Yukawa interaction is an exception. We omit the corresponding dilaton exponential in the Yukawa interaction in Eq.~\eqref{eq:Smin} to avoid generating a fermion mass in flat backgrounds. We discuss this further in Sec.~\ref{sec:bh}.} We impose $CP$ for simplicity. A complete list of independent on-shell three- and four-point amplitudes up to dimension eight, together with their correspondence to the operators in Eq.~\eqref{eq:LEFT}, is given in App.~\ref{app:eft}. As regards power counting, the coefficients of the four-boson operators scale as $\Lambda^{-4}$. Operators containing a fermion bilinear break chiral symmetry in the same way as the Yukawa interaction, which suggests $\cPF\sim\cDP\sim y\Lambda^{-3}$. At one loop, we keep terms of order $\kappa^4/16\pi^2$, $\kappa^2y^2/16\pi^2$ and $\kappa y\,\cPF/16\pi^2$, and neglect $\mathcal{O}(y^4/16\pi^2)$.\footnote{At this order, $y^4$ terms only enter the four-scalar amplitude through the fermion box.} Given the operators considered in Eq.~\eqref{eq:LEFT}, we truncate the EFT expansion of the amplitudes at $\mathcal{O}(E^4/\Lambda^4)$. Finally, we define the running of a coefficient $c_i$ at a scale $E<\Lambda$ as follows:
\begin{equation}
    \frac{dc_i(E^2)}{d\log E}=\gamma_i\,,
    \label{eq:running}
\end{equation}
so a positive $\gamma_i$ means that $c_i$ decreases towards the IR.

\section{Positivity at one loop}
\label{sec:positivity}

We can now study positivity bounds on the coefficients of the four-boson operators in Eq.~\eqref{eq:LEFT} obtained from consistency constraints on the $S$-matrix. For a $2\to2$ scattering amplitude $\mathcal{A}_4(s,t)$ analytic in the complex $s$ plane (except for poles and cuts on the real axis) at fixed $t<0$,\footnote{We use an all-incoming convention for the external legs in which $s=(p_1+p_2)^2$, and $t=(p_1+p_3)^2$.} and with an appropriate asymptotic behavior, Cauchy's theorem relates an integral of $\mathcal{A}_4$ over arcs of radius $\sim S<\Lambda^2$, computable within the EFT, to integrals of its discontinuities from $S$ to infinity. See Fig.~\ref{fig:contour} and App.~\ref{app:bounds} for more details. We call the IR side of these dispersion relations the \emph{arc} $a(S)$. For elastic amplitudes or suitable combinations with other amplitudes, unitarity makes the discontinuities positive in the forward limit. As a consequence, we find: first, $a(S)>0$, i.e., \emph{positivity}; and second, $da/dS=-\rho(S)/S^3\le0$, where $\rho(S)\ge0$ is the spectral density at $s=S$, i.e., the arc increases monotonically towards the IR. Moreover, polynomial (tree-level) terms in the amplitude beyond $s^2$ do not contribute to the arcs that we consider in this work (since we target the dimension-eight operators of Eq.~\eqref{eq:LEFT} with our dispersion relations), so the whole $S$ dependence of $a(S)$ for $S<\Lambda^2$ comes from loops of the light states. Due to monotonicity, the strongest bound on $a(S)$ is the one at the largest $S$ where the EFT is under control~\cite{Bellazzini5}. Bounds at lower $S$ follow from it. Let us now consider each operator in turn.

\subsection{Four-photon bounds}
\label{sec:F4}

At tree level, the dispersion relation for the elastic amplitude $\mathcal{A}_4(1_{\gamma^+},2_{\gamma^-},3_{\gamma^-},4_{\gamma^+})$ in the forward limit gives
\begin{equation}
    8\cFa+8\cFb=\lim_{t\to0}\frac1\pi\int_S^\infty\frac{ds'}{(s')^3}\left[{\rm Im}\mathcal{A}_4(1_{\gamma^+},2_{\gamma^-},3_{\gamma^-},4_{\gamma^+})+{\rm Im}\mathcal{A}_4(1_{\gamma^+},2_{\gamma^+},3_{\gamma^-},4_{\gamma^-})\right]>0\,,
    \label{eq:F4tree}
\end{equation}
and suitable combinations of this relation with that for $\mathcal{A}_4(1_{\gamma^+},2_{\gamma^+},3_{\gamma^+},4_{\gamma^+})$ bound each coefficient separately~\cite{Hamada1}:
\begin{equation}
    \cFa>0\,,\qquad\cFb>0\,.
    \label{eq:F4treesep}
\end{equation}
Without further corrections, these results lead to the claim that causality and unitarity enforce the WGC in Einstein--Maxwell theory~\cite{Hamada1,Bellazzini4}.

We can improve the accuracy of the bounds by including one-loop contributions. In practice, this amounts to correcting the IR integral by using the one-loop amplitude expression; positivity of the UV side is an all-order statement. Starting with $\cFb$ and working up to $\mathcal{O}(s^2/\Lambda^4)$, its bound becomes
\begin{equation}
    \cFb(S)>-\frac{\kappa^4}{512\pi^2}\,,\qquad\gFb=-\frac{\kappa^4}{96\pi^2}\,.
    \label{eq:F4b}
\end{equation}
The constant term $\kappa^4/512\pi^2$ is a contribution to the arc that arises from rational terms of the one-loop amplitude.\footnote{One-loop amplitudes employed in this section can be found in App.~\ref{app:oneloop}.} Since $\gFb$ is negative, $\cFb$ grows in the IR, and the bound is automatically satisfied below the matching scale if it is satisfied there.

\begin{figure}[t]
    \centering
    \includegraphics[width=1.0\textwidth]{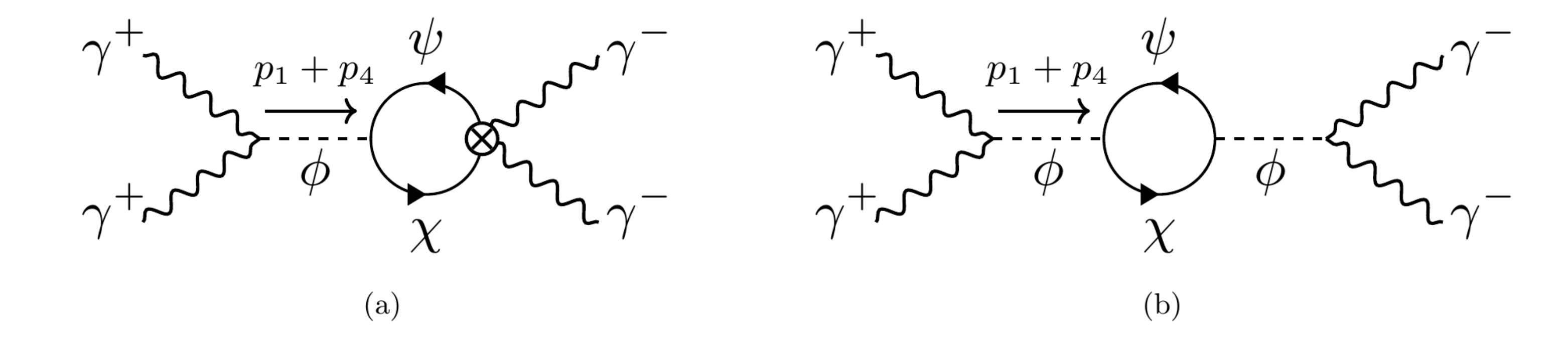}
    \caption{(a) Scalar exchange dressed by a fermion bubble with an insertion of $\cPF$, which renormalizes $\cFa$. (b) The same topology without the insertion, which produces the $1/S$ term in Eq.~\eqref{eq:F4a}. Neither diagram has a $t$-channel pole, so the forward limit can be taken.}
    \label{fig:diagrams}
\end{figure}

The bound on $\cFa$ is more interesting:
\begin{equation}
    \cFa(S)>-\frac{\kappa^2y^2}{64\pi^2S}-\frac{\kappa^4}{512\pi^2}\,,\qquad\gFa=-\frac{\kappa^4}{96\pi^2}+\frac{\cPF\,\kappa y}{8\pi^2}\,,
    \label{eq:F4a}
\end{equation}
where we have applied the same $\mathcal{O}(s^2/\Lambda^4)$ EFT truncation. In this case, two features are new compared with the tree-level bound and with the purely bosonic theory. The first one concerns the sign of the anomalous dimension $\gFa$, which receives a contribution from the diagram of Fig.~\ref{fig:diagrams}(a). It is linear in the Yukawa, in $\kappa$, and in the coefficient $\cPF$ of Eq.~\eqref{eq:LEFT}, whose sign is not fixed by any bound in the theory. Then, for
\begin{equation}
    \cPF\,\kappa y>\frac{\kappa^4}{12}
    \label{eq:negrun}
\end{equation}
we have $\gFa>0$. In this scenario, the coefficient $\cFa$ decreases logarithmically towards the IR and eventually becomes negative. Remarkably, this is an instance of negative running of $\cFa$ which is not caused by gravitational effects, so it is unsuppressed by $\MPl$, in contrast to the mechanisms proposed in~\cite{Charles1,ArkaniHamed2}. Concerning power counting, with $\cPF\sim y\Lambda^{-3}$, Eq.~\eqref{eq:negrun} requires $y^2\gtrsim(\kappa\Lambda)^3$, which is still compatible with having negligible $\mathcal{O}(y^4)$ contributions.

\begin{figure}[t]
    \centering
    \includegraphics[width=0.49\textwidth]{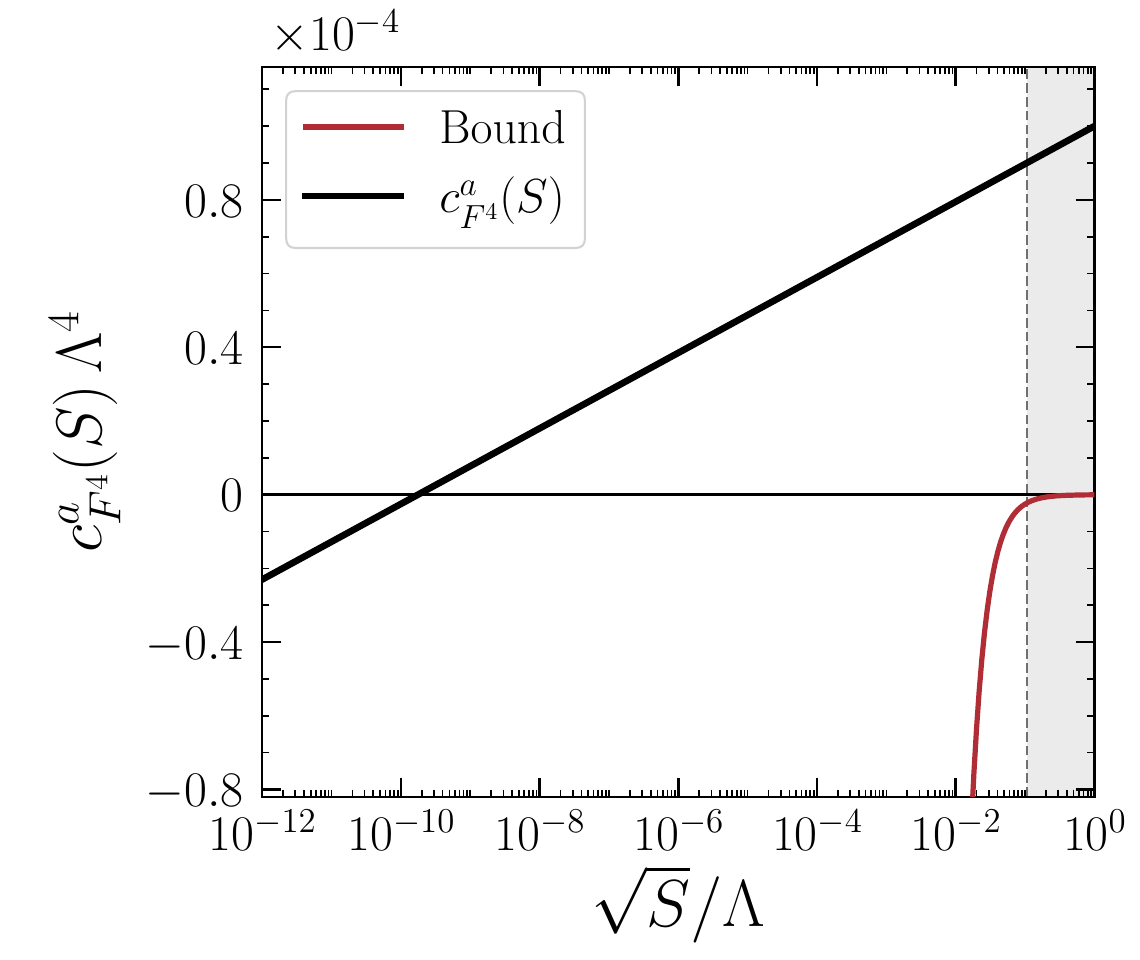}\hfill
    \includegraphics[width=0.49\textwidth]{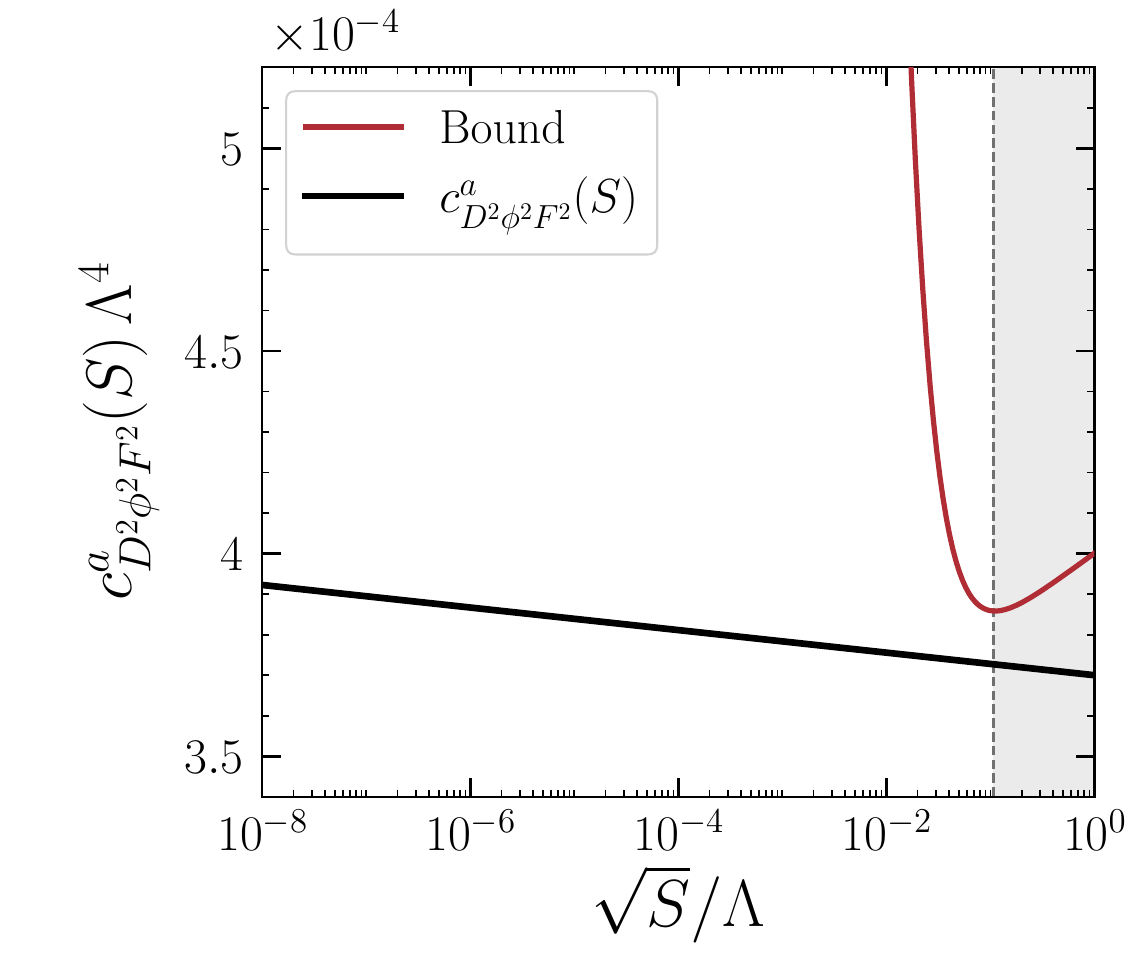}
    \caption{Running of (a) $\cFa(S)$ and (b) $\cDFa(S)$ (black), together with the four-photon lower bound of Eq.~\eqref{eq:F4a} and the mixed upper bound of Eq.~\eqref{eq:DFinel1} (red). We use $\kappa(\Lambda)=0.1\,\Lambda^{-1}$, $y=0.04$, $\cPF=0.09\,\Lambda^{-3}$, $\cDP=-0.02\,\Lambda^{-3}$ and $\cFa(\Lambda^2)=10^{-4}\Lambda^{-4},\,\cDFa(\Lambda^2)=3.7\times10^{-4}\Lambda^{-4},\,\cD(\Lambda^2)=4\times10^{-4}\Lambda^{-4}$. We set $\cDFb(\Lambda^2)=0$ for reasons explained in the following section. The coupled RG equations include the running of $\kappa$ in Eq.~\eqref{eq:kapparun}. The grey region $\sqrt{S}>\sqrt{S_*}\simeq0.107\Lambda$ indicates where the truncated four-photon arc fails the monotonicity test, so the displayed bounds are unreliable. $\cFa$ can become negative at sufficiently low energies, while $\cDFa$ can be positive and grow towards the IR.}
    \label{fig:running}
\end{figure}

The second new feature is the relaxation of the bound produced by the fermion bubble in scalar exchange, Fig.~\ref{fig:diagrams}(b), which adds the term $-\kappa^2y^2/(64\pi^2S)$. Since this term grows as $1/S$ into the IR while $\cFa(S)$ decreases only logarithmically, a coefficient that satisfies Eq.~\eqref{eq:F4a} near the matching scale satisfies it at all lower scales, however negative it becomes. Fig.~\ref{fig:running}(a) shows an example. The scalar therefore plays a role analogous to graviton exchange in Refs.~\cite{Henriksson1,Chang1}, but through a coupling that is not suppressed by $\MPl$.  Thus, $\cFa(S)$ can become arbitrarily large and negative in the IR while staying consistent with causality and unitarity.

Notice that the Yukawa interaction also induces a running of $\kappa$ through the wave-function renormalization of the scalar. This can be read directly from the logarithmic correction to the $u$-channel scalar exchange in the one-loop four-photon amplitude (see Eq.~\eqref{eq:C3}), yielding
\begin{equation}
    \frac{d\kappa^2}{d\log E}=\frac{y^2\kappa^2}{4\pi^2}\,.
    \label{eq:kapparun}
\end{equation}
Thus, $\kappa$ decreases towards the IR. We take into account this running when solving coupled RG evolution equations.

Before we move to the analysis of other operators, we recall that the bounds shown here, especially Eq.~\eqref{eq:F4a}, are subject to the assumptions of EFT truncation, i.e., they only hold for energy scales $S\lesssim\Lambda^2$. Monotonicity offers a particularly simple way to evaluate the range of validity of the truncated expansion. If we require monotonic decrease of the arc leading to Eq.~\eqref{eq:F4a},
\begin{equation}
    \frac{da(S)}{dS}\leq0,\qquad a(S)=\cFa(S)+\kappa^2y^2/(64\pi^2S)+\kappa^4/(512\pi^2),
    \label{eq:monotonicarc}
\end{equation}
we find that this is true for
\begin{equation}
    S<S_*\equiv\frac{\kappa^2y^2}{32\pi^2\gFa}\,.
    \label{eq:Sstar}
\end{equation}
Above $S_*$, the truncated arc increases with $S$, whereas the exact arc cannot. Therefore, the truncated expression must fail there. We take $S_*$ as an approximate upper limit on the validity of Eq.~\eqref{eq:F4a}.\footnote{Nevertheless, notice that $S<S_*$ is only a necessary condition for consistency of the truncation.} For the illustrative benchmark in Fig.~\ref{fig:running}, $S_*\simeq0.01\Lambda^2$.\footnote{In the case of Eq.~\eqref{eq:F4a}, the $1/S$ term and the running of $\cFa$ share a common origin in the dispersive representation. In the EFT region $S\le s\le\Lambda^2$, the relevant intermediate state is the fermion pair $\psi\chi$, and the corresponding contribution to the spectral density in the dispersive sum for $\cFa$ is proportional to the modulus squared of the amplitude $\gamma^+\gamma^+\to\psi\chi$. Unitarity therefore gives the positive spectral density:
\begin{equation}
    \frac{\rho_{\psi\chi}(s)}{s^3}=\frac{1}{64\pi^2}\left(\frac{\kappa y}{s}-2\cPF\right)^2\ge0\,.
    \label{eq:square}
\end{equation}
Expanding the square, the $\kappa^2 y^2/s^2$ term produces the $1/S$ term of the arc, and the cross term produces the running of $\cFa$. Due to EFT truncation, we have thus discarded the last term $\propto \cPF^2$ which would restore monotonicity. Still, this term represents only one missing contribution to the arc at energies close to $\Lambda$. There are other terms renormalizing higher-order operators which would enter the arc if truncation were not applied.}

\subsection{Four-scalar bound}
\label{sec:phi4}

At tree level, it is well known that $\cD>0$~\cite{Adams1}. At one loop and in our massless theory, the Yukawa interaction generates collinear IR divergences in amplitudes with external scalars, which complicate the positivity analysis. These divergences arise from massless fermion bubbles on the external legs and are proportional to the tree-level amplitude, with no additional kinematic dependence:
\begin{equation}
    \mathcal{A}_4(1_\phi,2_\phi,3_\phi,4_\phi)\big|_{\rm IR\,div.}=\mathcal{A}_4\big|_{\rm tree}\times\frac{-y^2}{4\pi^2}\log\left(\frac{\mu^2}{m_{\rm IR}^2}\right),\qquad\mathcal{A}_4\big|_{\rm tree}=2\cD(s^2+t^2+u^2)\,,
    \label{eq:IRdiv}
\end{equation}
with $m_{\rm IR}\to0$ an IR regulator and $\mu$ the renormalization scale. As a consequence, IR divergences enter the one-loop arcs. Nevertheless, their origin in massless bubbles on external legs and their kinematic dependence suggest that they should factorize in the same way on both sides of the dispersion relation~\cite{Gervais1,Baratella2}. Assuming such a cancellation, we obtain an IR-safe bound
\begin{equation}
    \cD(S)>0\,,\qquad\gamma_{D^4\phi^4}=-\frac{\kappa^4}{16\pi^2}+\frac{y^2}{2\pi^2}\cD\,.
    \label{eq:phi4}
\end{equation}
The factorization of these IR divergences in the dispersive argument is an explicit assumption of this work.

\subsection{Two-scalar--two-photon bounds}
\label{sec:DF}

The operators $c^{a,b}_{D^2\phi^2F^2}$ contribute to $\mathcal{A}_4(1_{\gamma^+},2_\phi,3_{\gamma^-},4_\phi)$ and to $\mathcal{A}_4(1_{\gamma^+},2_{\gamma^+},3_\phi,4_\phi)$. The former is elastic, and its contact contribution only depends on $\cDFb$. Including loop corrections, with the IR divergences treated as above,
\begin{equation}
    \cDFb(S)<0\,,\qquad\gamma^b_{D^2\phi^2F^2}=\frac{\kappa^4}{12\pi^2}+\frac{y^2}{4\pi^2}\cDFb\,.
    \label{eq:DFb}
\end{equation}
The negative sign is a consequence of the forward limit of the polarizations, $\mathcal{A}_4(1_{\gamma^+},2_\phi,3_{\gamma^-},4_\phi)\sim-s^2\cDFb$. For $\cDFa$, there is no elastic bound, but two-sided inelastic bounds follow from Cauchy--Schwarz applied to the dispersive sums~\cite{Trott1} (see App.~\ref{app:bounds}). At tree level, these bounds read
\begin{equation}
    -2\sqrt{\cD\,\cFa}\le\cDFa\le2\sqrt{\cD\,\cFa}-\frac12\cDFb\,.
    \label{eq:DFinel0}
\end{equation}
The upper bound is the relevant one for our discussion in the following section. At one loop, it is modified in the following way:
\begin{equation}
    \cDFa(S)\le2\sqrt{\cD(S)\left(\cFa(S)+\frac{\kappa^2y^2}{64\pi^2S}+\frac{\kappa^4}{512\pi^2}\right)}-\frac12\cDFb(S)\,,
    \label{eq:DFinel1}
\end{equation}
with
\begin{equation}
    \gamma^a_{D^2\phi^2F^2}=-\frac{\kappa^4}{48\pi^2}+\frac{\cDP\,\kappa y}{8\pi^2}+\frac{y^2}{4\pi^2}\cDFa\,.
    \label{eq:DFinel2}
\end{equation}
The coefficient $\cDP$ renormalizes $\cDFa$ in the same way that $\cPF$ renormalizes $\cFa$, so $\gamma^a_{D^2\phi^2F^2}$ is also sign-indefinite. Notice that the upper bound of Eq.~\eqref{eq:DFinel1} allows $\cDFa$ to be positive and increase towards the IR, as illustrated in Fig.~\ref{fig:running}(b).

To conclude, we summarize the main results of our analysis of the positivity bounds in Eqs.~\eqref{eq:F4b}, \eqref{eq:F4a}, \eqref{eq:phi4}, \eqref{eq:DFb} and \eqref{eq:DFinel1}, which reveal two important qualitative differences from the tree-level bounds and the purely bosonic theory. First, interference between Yukawa interactions and fermionic EFT operators renders the anomalous dimensions of $\cFa$ and $\cDFa$ sign-indefinite. Second, the fermion bubble in scalar exchange relaxes the bound on $\cFa$ by a term proportional to $1/S$, allowing the coefficient to take increasingly negative values towards the IR without violating positivity. Analogously, this same term relaxes the bound on $\cDFa$ in the IR. In the next sections, we discuss some of the relevant phenomenological consequences of these features.

\section{Implications for extremal black holes in EMd theory}
\label{sec:bh}

Given the relevance of the sign of four-photon operators for the WGC, the natural question to ask is whether our findings about consistent negative running can have any effect on the decay of extremal BHs. Classically, BH solutions corresponding to Eq.~\eqref{eq:Smin} are those of the EMd theory~\cite{Gibbons1,Garfinkle1,Cheng1,Poletti1}, if radiative corrections to the scalar potential are negligible (which is a nontrivial assumption, as discussed below). In~\cite{Loges1}, the four-derivative operator corrections to the extremal mass have already been computed (at linear order in WCs), so one could be tempted to evaluate the RG-improved charge-to-mass ratio and determine whether extremality is prevented by negative running. We devote the following discussion to studying this argument. After that, we will see how a more cautious approach reveals important obstructions to it.

\subsection{RG Evolution effects}

Let us study generic dyonic EMd BHs.\footnote{For details about the BH solution, see App.~\ref{app:profiles}.} For the WGC, we are interested in the extremal solutions, which have horizon area and horizon value of the dilaton~\cite{Loges1}
\begin{equation}
    \lim_{T\to0}A_+=4\pi|q_e||q_m|\,,\qquad\lim_{T\to0}e^{-2\kappa\phi}\big|_{r=r_H}=\frac{|q_e|}{|q_m|}\equiv\frac1\xi\,.
    \label{eq:attractor}
\end{equation}
Notice that this defines the characteristic size $L=\sqrt{A_+/4\pi}=\sqrt{|q_e||q_m|}$ and the corresponding energy scale $1/L$. At two derivatives and for $\phi_\infty=0$, the extremality condition, i.e., the extremal charge-to-mass ratio, reads~\cite{Cremonini1}
\begin{equation}
    \frac{8\pi\sqrt{|q_e||q_m|}}{M}\,g_\kappa\!\left(\tfrac12\log\xi\right)\Big|_{T=0}=1\,,
    \label{eq:ext2der}
\end{equation}
where $g_\kappa(\eta)$ is a function defined by the following equation and initial condition:
\begin{equation}
    g_\kappa(\eta)^2+2\kappa^2\left(\frac{dg_\kappa}{d\eta}\right)^2=\cosh(2\eta)\,,\quad g_\kappa(0)=1\,.
\end{equation}
Now, the operators of Eq.~\eqref{eq:LEFT} correct the right-hand side by terms of order $c_i/L^2$~\cite{Loges1,Reall1}. Thus, the presence of superextremal states is determined by the sign of these corrections. For instance, at $\xi=1$, $|q_e|=|q_m|=q$, only $\cFb$ contributes, and the extremality condition reads
\begin{equation}
    \frac{8\pi q}{M}=1+\frac{8\cFb}{5q^2}+\dots\, ,
\end{equation}
for any $\kappa$, so positivity of $\cFb$ implies superextremality there. Given the result of Eq.~\eqref{eq:F4b}, superextremality is guaranteed except for a possible loop-suppressed negative region for the smallest BHs within the EFT (i.e., for those with horizon size $1/L\sim\Lambda$).

In setups with multiple charges (magnetic and electric in our case), the WGC is generalized into the convex-hull condition (CHC)~\cite{Cheung2}. In essence, the CHC states that, for every extremal BH to be able to decay into lighter states, the convex hull defined by the positions of the lighter states of mass $m$ in the $(q_e/m, q_m/m)$ plane must contain the location of the extremal BH in that plane. Because the extremality curve of EMd BHs develops increasingly sharp and extended tips along the electric and magnetic axes as $\kappa$ grows~\cite{Cremonini1}, for $\kappa\gg1$ the CHC~\cite{Cheung2} depends on the presence or absence of superextremal states near the magnetic and electric limits.\footnote{This means that superextremality at $\xi=1$ is not relevant for the decay of purely magnetic and electric BHs.} Since the electric limit is singular~\cite{Cremonini1}, we focus instead on the magnetic one, which, for $\kappa\gg1$, features a corrected extremality condition given by~\cite{Loges1,ArkaniHamed2}
\begin{equation}
    \frac{4\pi|q_m|}{M}\sqrt{\frac{2}{1+2\kappa^2}}\,\bigg|_{\rm ext}=1+\frac{1}{5q_m^2\kappa^4}\left[4\cFa-2\cDFa+\cD\right]+\dots\,.
    \label{eq:ext_mag}
\end{equation}

Our first argument then consists of inserting the vacuum RG-evolved coefficients at $E\sim1/L$ into Eq.~\eqref{eq:ext_mag} and examining whether the positivity constraints presented in the previous section imply superextremality. A first difficulty arises from the fact that extremal purely magnetic BHs have vanishing horizon area, so their near-horizon region lies beyond the reach of our EFT. Nevertheless, dyonic BHs near the magnetic limit, which exhibit finite horizon area, can be assumed to feature a charge-to-mass ratio that approaches the purely magnetic result as $\xi=|q_m/q_e|\to\infty$. This correspondence holds at the two-derivative level for arbitrary $\kappa$, and has also been verified for the leading four-derivative corrections in cases where an analytic solution is known~\cite{Loges1}. This motivates considering sufficiently magnetic dyonic BHs, with $\Lambda L\gg1$, as the setting in which we can test the WGC using Eq.~\eqref{eq:ext_mag}.

\begin{figure}[t]
    \centering
    \includegraphics[width=0.50\textwidth]{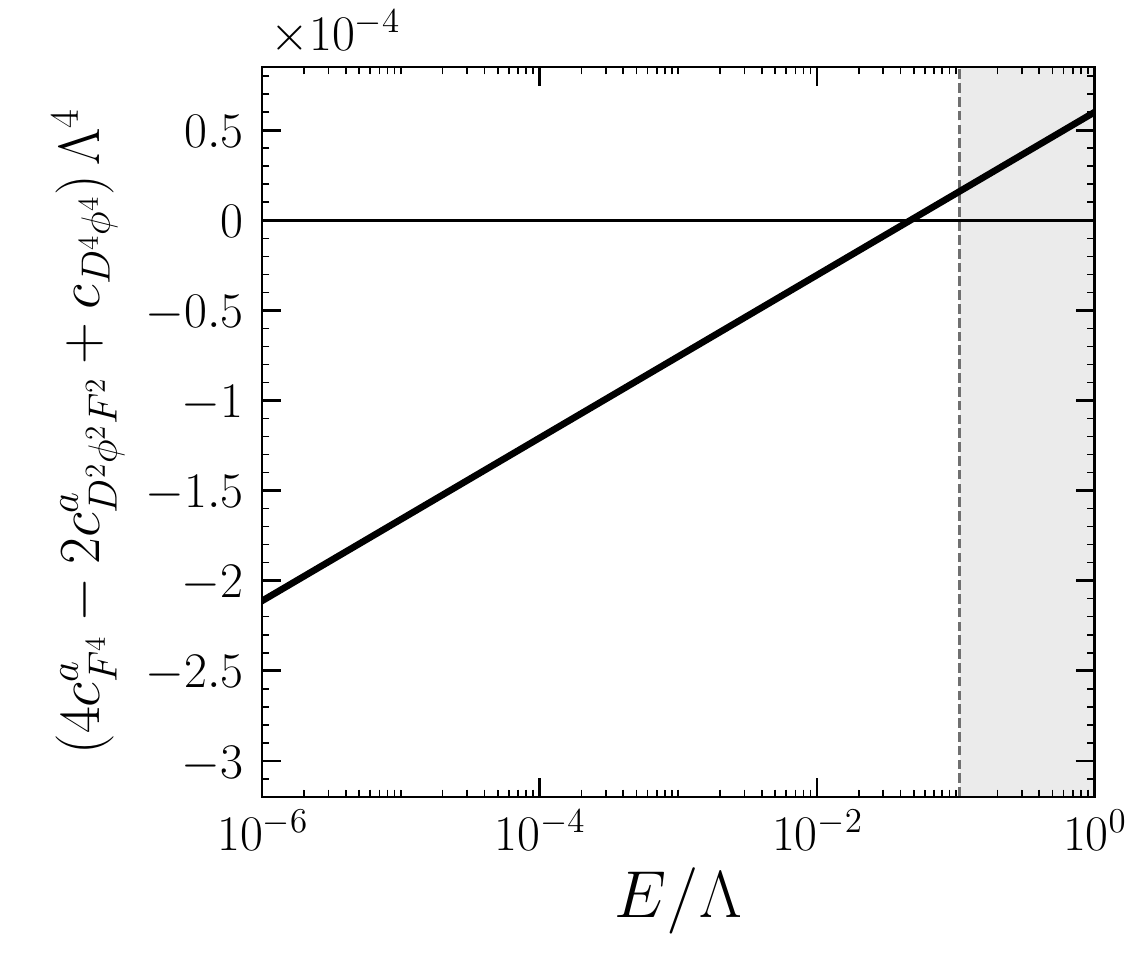}
    \caption{Running of the magnetic extremality combination $4\cFa-2\cDFa+\cD$ for the benchmark of Fig.~\ref{fig:running}. The grey region $E>\sqrt{S_*}$ marks where the truncated four-photon arc fails the monotonicity test.}
    \label{fig:combination}
\end{figure}

Under this assumption, the sign of the \emph{full} combination of WCs entering Eq.~\eqref{eq:ext_mag}, $4\cFa-2\cDFa+\cD$, determines the direction of the correction to magnetic extremality. For the benchmark of Fig.~\ref{fig:running}, which satisfies the displayed truncated bounds within their stated range of validity, Fig.~\ref{fig:combination} shows that this combination can become negative in the IR. In such an approximation, this pushes the charge-to-mass ratio towards the subextremal region.

Various comments are in order concerning this result. First, notice that given the tree-level constraints $\cFa,\cD>0$ and Eq.~\eqref{eq:DFinel0}, the combination need not be positive even at tree level. The fact that general tree-level positivity bounds need not guarantee a positive correction to magnetic extremality in EMd was already emphasized in~\cite{Loges1}. More explicitly, we find
\begin{equation}
    4\cFa-2\cDFa+\cD\ge\left(2\sqrt{\cFa}-\sqrt{\cD^{\vphantom{a}}}\right)^2+\cDFb,
\end{equation}
and $\cDFb<0$. In~\cite{ArkaniHamed2}, the authors argue that the leading threshold corrections to $4\cFa-2\cDFa+\cD$ from integrating out massive states in weakly coupled theories can be expressed as sums of squares on BH backgrounds, using the generalized optical theorem. This is explained by the fact that in this explicit construction the mixed photon--scalar channel is absent, so $\cDFb=0$ and the combination of WCs is indeed bounded by a perfect square.\footnote{Unlike the scalar-exchange example of~\cite{ArkaniHamed2}, heavy-vector exchange in the $\phi\gamma$ channel, heavy spin-2 exchange, or loops of charged states can generate nonzero $\cDFb$. In those cases, the matching value is negative, as required by positivity bounds.} The novelty of our result is that, even starting from matching conditions that enforce this positivity, the sign-indefinite running induced by the Yukawa interactions can drive the magnetic combination negative in the IR, without requiring the exponentially large scales and associated tuning encountered in the scenarios of Ref.~\cite{ArkaniHamed2}. Our benchmark explicitly uses $\cDFb=0$.

Second, the absence of superextremal states at lower energies does not, by itself, obstruct the CHC; we are completely agnostic about states beyond $\Lambda$. Therefore, even if we accept that the combination can be negative for all energies below $\Lambda$, we cannot claim that the WGC fails for this system.

Third, in obtaining Fig.~\ref{fig:combination}, we have included the running of $\kappa$ but neglected that of the Yukawa coupling. In the renormalizable scalar-fermion sector, the latter satisfies $dy/d\log E=5y^3/(16\pi^2)$ at one loop. Given the small value $y(\Lambda)=0.04$ of our benchmark, its relative variation down to the scale where the magnetic combination changes sign is approximately $1.55\times10^{-4}$. We have explicitly checked that including this running in the RG equations shifts the zero of the combination by less than $0.1\%$, without altering our conclusion. This provides a consistency check of the fixed-$y$ approximation, although a complete treatment of higher-order RG effects would also require accounting for the running of the remaining fermionic WCs.

Lastly, we recall that this argument based on RG evolution does not take into account the full one-loop expression for the charge-to-mass ratio. The substitution of $E\sim1/L$ in the RG-evolved WCs is indeed naive, and there might be further contributions to the extremality condition that we have not considered.

Even setting aside these qualifications, however, the argument faces a more fundamental obstruction arising from the physics of the BH backgrounds themselves: the same interaction responsible for the potentially dangerous running also gives rise to a dynamical effect that obstructs its contribution to the extremality condition. Rather than merely exposing a limitation of the RG-based argument, this reveals a new mechanism by which the BH background frustrates the apparent challenge to the WGC, to which we turn now.

\subsection{Black-hole background effects}
\label{sec:bhbackground}

From Eq.~\eqref{eq:attractor}, we obtain that the attractor value of our solution is $\phi_H=\log(\xi)/(2\kappa)$. For $\kappa\gg1$, the magnetic regime in Eq.~\eqref{eq:ext2der}, which corresponds to $g_\kappa\simeq e^\eta/(2\kappa)$, is obtained only if $e^\eta\gg2\kappa$, where $\eta=\frac12\log\xi$. Therefore, the attractor value of the scalar is $\phi_H\gtrsim\log(2\kappa)/\kappa$. Notice that this means that, through the Yukawa coupling of Eq.~\eqref{eq:Smin}, the fermion acquires a mass near the attractor, which, in units of $\Lambda$, is
\begin{equation}
    \frac{m_H}{\Lambda}=\frac{|y \phi_H|}{\Lambda}\gtrsim\frac{|y|}{\kappa\Lambda}\log\left(2\kappa\right),
\end{equation}
which becomes increasingly large for a large scale separation between $\Lambda$ and $\MPl$. Our first argument relied on light-fermion loops contributing to the running of WCs entering the charge-to-mass ratio. If the fermion has a mass much larger than $\Lambda$, it can decouple at energies higher than those of interest, and its contribution to the running freezes. This is important because if the fermion does not enter the running of the WCs, the only contribution to the renormalization of such terms comes from dilaton couplings and
\begin{equation}
    \frac{d\left( 4\cFa-2\cDFa+\cD\right)}{d\log E}=-\frac{\kappa^4}{16\pi^2},
\end{equation}
so the combination starts growing in the IR.

Nevertheless, the fact that the fermion becomes so massive near the horizon does not automatically mean that the effects of (negative) running completely disappear. Critically, the fermion eventually becomes effectively massless as the distance from the horizon increases and the electromagnetic invariant $F^2$ vanishes precisely at the horizon. Indeed, for our dyonic solutions,
\begin{equation}
    F^2=\frac{2}{R^4}\left(q_m^2-q_e^2e^{4\kappa\phi}\right),
    \label{eq:EMdyonic}
\end{equation}
where $R$ is the areal radius. Using the attractor condition $e^{2\kappa\phi_H}=\xi=|q_m/q_e|$, we obtain
\begin{equation}
    \left.F^2\right|_{r=r_H}
    =\frac{2}{R_H^4}\left(q_m^2-q_e^2\xi^2\right)=0.
\end{equation}
Consequently, the contributions of the operators proportional to $\cFa$ and $\cDFa$ vanish locally at the horizon, precisely where the fermion acquires its largest mass. Therefore, to assess the extent to which the induced fermion mass obstructs our previous argument, a more careful analysis is required.

Indeed, one can consider the corrections to the extremality condition as the result of integrating a weight associated with each dimension-eight operator over the entire radial profile, where the fermion mass varies. Thus, the correct analysis consists of identifying the region that contributes the most to the corrections in Eq.~\eqref{eq:ext_mag} and evaluating the mass of the fermion there. Following the procedure of~\cite{Loges1}, the local corrections are obtained by integrating the dimension-eight operators over the uncorrected BH background. For simplicity, let us first work with the purely magnetic background. After substituting the scalar and electromagnetic profiles and integrating over the angles, we are left with the following radial integrals (with the WCs omitted):\footnote{More details about the solutions and the procedure can be found in App.~\ref{app:profiles}.}
\begin{subequations}
    \label{eq:magneticspatialmain}
    \begin{align}
        4\pi\int_{r_0}^{\infty}\!dr\,R^2e^{-6\kappa\phi}(F_{\mu\nu}F^{\mu\nu})^2
        &=16\pi q_m^4\int_{r_0}^{\infty}\!\frac{dr}{r^6}\,,\\
        4\pi\int_{r_0}^{\infty}\!dr\,R^2e^{-4\kappa\phi}\big(\partial_\mu\phi\partial^\mu\phi\big)F_{\nu\rho}F^{\nu\rho}
        &=-8\pi q_m^2\left(\frac{2\kappa r_0}{1+2\kappa^2}\right)^2
        \int_{r_0}^{\infty}\!\frac{dr}{r^6}\,,\\
        4\pi\int_{r_0}^{\infty}\!dr\,R^2e^{-2\kappa\phi}\big(\partial_\mu\phi\partial^\mu\phi\big)^2
        &=4\pi\left(\frac{2\kappa r_0}{1+2\kappa^2}\right)^4
        \int_{r_0}^{\infty}\!\frac{dr}{r^6}\,,
    \end{align}
\end{subequations}
where $r_0=|q_m|\sqrt{(1+2\kappa^2)/2}$ corresponds to the horizon location of the magnetic solution in this radial coordinate. We observe that the three operators contribute with the same radial dependence, $dr/r^6$. Using the thermodynamic relation of~\cite{Loges1}, the correction to the extremality condition can be expressed, for $\kappa\gg1$, as
\begin{equation}
    \frac{4\pi|q_m|}{M}\sqrt{\frac{2}{1+2\kappa^2}}\,\bigg|_{\rm ext}-1
    =\frac{1}{q_m^2\kappa^4}\int_{r_0}^{\infty}\!dr\,\frac{r_0^5}{r^6}
    \left[4\cFa-2\cDFa+\cD\right]+\dots\,,
    \label{eq:magneticradial}
\end{equation}
which reproduces Eq.~\eqref{eq:ext_mag} upon integration. We have therefore found the radial distribution of the corrections.

\begin{figure}[t]
    \centering
    \includegraphics[width=0.70\textwidth]{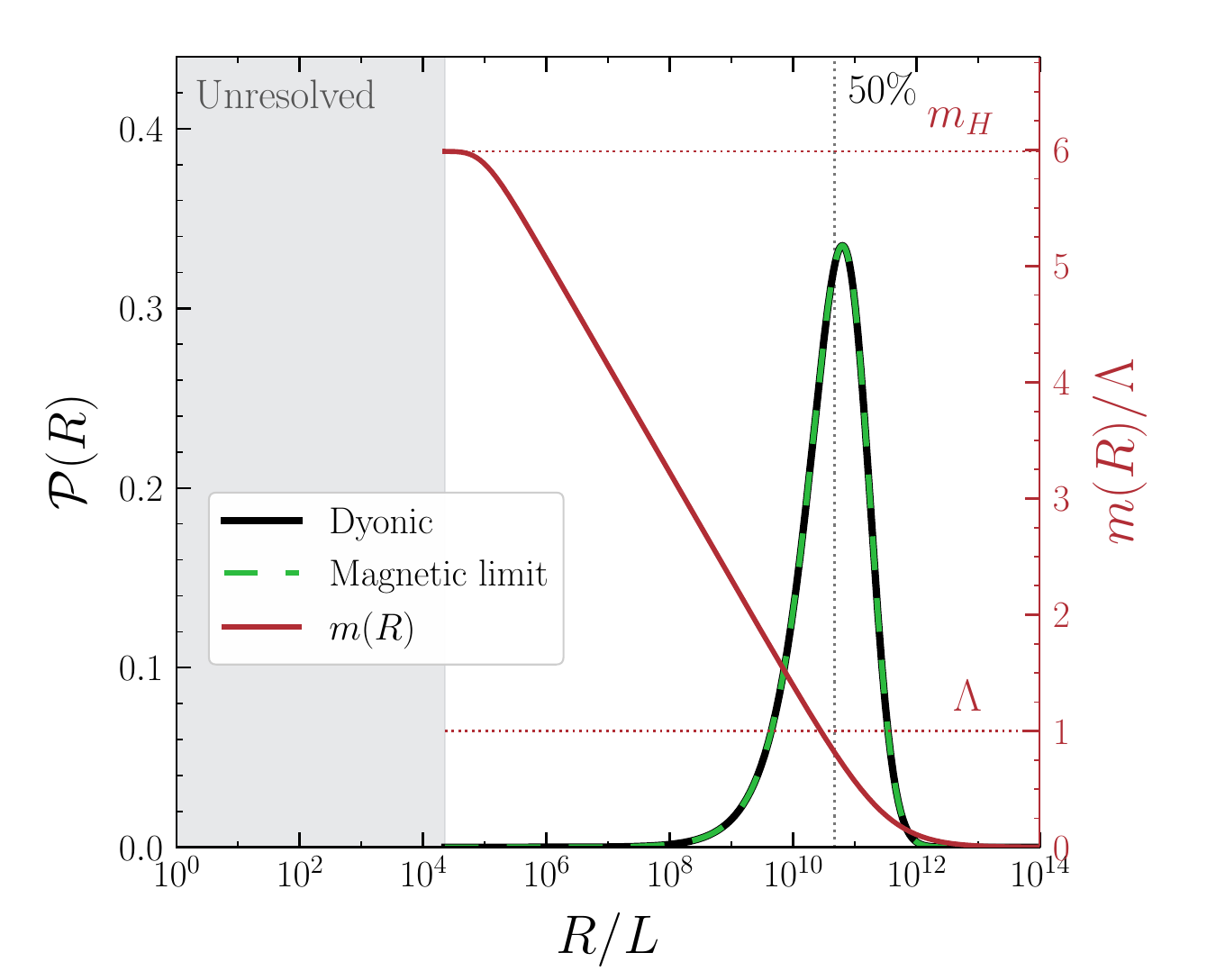}
    \caption{Normalized radial weight of the four-derivative contributions to the extremality condition $\mathcal{P}(R)$ (left axis) and the induced fermion mass $m(R)/\Lambda$ (right axis) versus the areal radius $R/L$. The black curve is the normalized dyonic density for the $F^4$ operator, acting as a representative of the three relevant operators, and the green dashed curve is their common magnetic-limit distribution. The red curve shows the dyonic fermion mass. We used $\kappa=10^5$, $\xi=10^{13}$, $\kappa\Lambda=0.1$, $y=0.04$. The vertical dotted line marks the radius at which $50\%$ of the weight has accumulated; the red dotted lines mark $m=\Lambda$ and $m=m_H$. The shaded ``Unresolved'' region denotes the range not covered by the direct dyonic numerical integration.}
    \label{fig:F4weight}
\end{figure}

It turns out that this purely magnetic distribution accurately approximates the normalized radial distributions of all three local operators evaluated on the dyonic background in the relevant region, as explicitly shown in Fig.~\ref{fig:F4weight} for $F^4$. There we define a normalized distribution in the areal radius,
\begin{equation}
    \mathcal{P}(R)\,d\log(R/L)=5r_0^5\frac{dr}{r^6}\,,\qquad
    \mathcal{P}(R)\simeq\frac{5r_0^5R}{(R+r_0)^6}\, ,
    \label{eq:magneticweight}
\end{equation}
so $\mathcal{P}(R)$ measures the fraction of the total correction contributed per unit logarithmic interval in $R/L$. In Fig.~\ref{fig:F4weight} we therefore compare the result of the dyonic distribution (obtained numerically) with the known magnetic solution, and with the mass of the fermion $m(R)=|y\phi(R)|$ for the same benchmark of Fig.~\ref{fig:running} and in a BH with $\xi=10^{13}$. The conclusion is that most of the contribution to the extremality condition arises in a region where the mass of the fermion has dropped significantly compared with its value at the horizon, $m_H$, but remains sizeable compared with $\Lambda$, at least of order $\Lambda/10$. More concretely, at the half-weight radius, $m(R_{50})\simeq0.82\Lambda$, and it exceeds approximately $0.20\Lambda$ over the region containing the first $99\%$ of the weight. This is considerably larger than $0.047\Lambda$, the scale to which the magnetic combination was evolved in Fig.~\ref{fig:combination} to obtain a negative value. As a consequence, the background-induced mass obstructs the subextremality produced by negative running for our benchmark.

Now, to study whether this is the case more generally, we can evaluate $m(R_{50})$ in an arbitrary setup as a representative value of the mass of the fermion in the region where the corrections matter. In the purely magnetic approximation we can evaluate $m(R_{50})$ by using the formula of the areal radius in App.~\ref{app:profiles} at the radial coordinate within which 50\% of the weight accumulates, $r_{50}=2^{1/5}r_0$. We find
\begin{equation}
    \frac{m(R_{50})}{\Lambda}
    =\frac{2\kappa|y|}{(1+2\kappa^2)\Lambda}
    \log\!\left(\frac{1}{1-2^{-1/5}}\right)
    \simeq2\,\frac{|y|}{\kappa\Lambda},
    \label{eq:magneticmassmedian}
\end{equation}
for $\kappa\gg1$. Therefore, for perturbative but not excessively small parameters, we will find that the fermionic mass obstructs the RG-based argument for subextremality. Of course, we refer to scenarios in which, given $\mathcal{O}(\Lambda^{-4})$ WCs without fine-tuning at energies close to $\Lambda$, a negative combination is achieved through running. In those cases, we find that the fermion mass is too large for the coefficients to run far enough to enter the subextremal region. Even if we reduced the ratio $|y|/(\kappa\Lambda)$, the obstruction persists: since the negative running of $\cFa$ requires $y^2\gtrsim (\kappa \Lambda)^3/12$, the mass of the fermion satisfies
\begin{equation}
    \frac{m(R_{50})}{\Lambda}
    \gtrsim\, \sqrt{\kappa\Lambda}.
\end{equation}
Now, this means that the mass of the fermion can be reduced if we choose lower values of $\kappa\Lambda$. However, at the same time the WCs will run slower. In particular, for a natural setup, the energy at which the combination of WCs in the magnetic limit becomes negative, $E_0$, scales as
\begin{equation}
    \frac{E_0}{\Lambda}\sim e^{-\frac{\mathcal{O}(\Lambda^{-4})}{\gamma}},
\end{equation}
where $\gamma$ is a generic positive running coefficient of order $\mathcal{O}(y^2\kappa\Lambda^{-3})$. Hence, as $\kappa \Lambda$ decreases, the energy $E_0$ decreases faster than the scale at which the fermions decouple from the running, preventing the running mechanism from flipping the sign of the combination. In this regard, our benchmark is even favorable to the running mechanism; the WCs are individually small at the matching scale, so less running is required to drive the magnetic combination negative.

\subsection{Radiative stability of the EMd backgrounds}
\label{sec:scalarpotential}

As follows from Eq.~\eqref{eq:EMdyonic}, the electromagnetic field decreases with the areal radius, with $F^2\propto R^{-4}$ in the magnetic approximation. Given the large value of $R_{50}$ found above, the electromagnetic contribution to the dilaton equation is therefore strongly suppressed in the region where the corrections to extremality are concentrated. This raises the important consistency question of the radiative corrections to the scalar potential in this region. Indeed, the Yukawa interaction generates a Coleman--Weinberg potential which, for a Dirac fermion with $m(\phi)=|y\phi|$, reads~\cite{Coleman1,Martin1}
\begin{equation}
    V_{\rm CW}(\phi)=-\frac{m(\phi)^4}{16\pi^2}\left[\log\!\left(\frac{m(\phi)^2}{\mu^2}\right)-\frac32\right]
\end{equation}
in the $\overline{\rm MS}$ scheme. Including an effective scalar potential modifies the dilaton equation to
\begin{equation}
    \Box\phi=V'_{F^2}-V'_{\rm eff}(\phi)\,,\qquad V'_{F^2}\equiv\frac{\kappa}{2}e^{-2\kappa\phi}F^2\,.
    \label{eq:dilatoneq}
\end{equation}
We can estimate the relative importance of these contributions using our benchmark. Neglecting logarithms, and evaluating the fermionic contribution before possible cancellations with other terms in the renormalized potential, we find
\begin{equation}
    \left|\frac{V'_{\rm CW}}{V'_{F^2}}\right|\simeq\frac{y\,m(\phi)^3e^{2\kappa\phi}}{2\pi^2\kappa F^2}\,.
    \label{eq:ratio1}
\end{equation}
At $R_{50}/L\simeq4.70\times10^{10}$, this gives $|V'_{\rm CW}/V'_{F^2}|\sim 10^{17}(\Lambda L)^2$ for our benchmark. The fermionic contribution to the potential would therefore overwhelm the classical electromagnetic source if left uncompensated, invalidating the EMd approximation in the region relevant to our analysis. Although the numerical estimate depends on the renormalization conditions, the enormous hierarchy reveals that radiative stability requires a highly restrictive condition on the scalar potential.

One possibility to preserve the EMd background is to cancel the radiatively generated potential through additional degrees of freedom. For instance, consider four real neutral scalars $S_a$ with interactions
\begin{equation}
    \Delta\mathcal{L}=-\frac12y^2\phi^2\sum_{a=1}^{4}S_a^2\,.
    \label{eq:fourscalars}
\end{equation}
Their field-dependent masses satisfy $m_{S_a}^2(\phi)=y^2\phi^2=m(\phi)^2$, so their combined bosonic contribution exactly cancels the fermionic Coleman--Weinberg potential at one loop for arbitrary constant $\phi$~\cite{Martin1}. In theories with unbroken supersymmetry, analogous cancellations along a flat direction can persist to all orders in perturbation theory~\cite{Seiberg1}. Importantly, these additional scalars have no direct couplings to photons or fermions. Their contributions to the amplitudes studied in the previous section first appear at $\mathcal{O}(y^4)$, which we neglect in our calculation. Thus, this extension removes the leading radiative contribution to the scalar potential without modifying the amplitude results or RG equations used in our analysis.

However, in general this is not sufficient. In particular, the one-loop cancellation does not automatically eliminate higher-loop contributions, which can remain much larger than the electromagnetic source despite their additional loop suppression. Neglecting logarithms and order-one coefficients, we estimate the two-loop contribution to the potential as
\begin{equation}
    V_{(2)}\sim\frac{m(\phi)^4}{16\pi^2}\,\frac{y^2}{16\pi^2}\,,
    \label{eq:V2loop}
\end{equation}
since the relation between the coupling in Eq.~\eqref{eq:fourscalars} and $y^2$ is not preserved by the RG evolution, and the two-loop vacuum diagrams do not cancel~\cite{Martin1}. The corresponding ratio $|V'_{(2)}/V'_{F^2}|$ is therefore suppressed relative to the one-loop estimate in Eq.~\eqref{eq:ratio1} by $y^2/(16\pi^2)\simeq10^{-5}$, giving $|V'_{(2)}/V'_{F^2}|\sim10^{12}(\Lambda L)^2$ for our benchmark.

We must therefore focus on theories in which the full renormalized effective potential is sufficiently flat over $0\leq\phi\leq\phi_H$. More precisely, its derivative must remain negligible compared with the classical sources in the region of interest, while its contribution to the Einstein equations must also be negligible. For example, at $R_{50}$ the former requires $|V_{\rm eff}'(\phi(R_{50}))|\ll3.43\times10^{-21}\Lambda^3/(\Lambda L)^2$. Preserving the original dyonic attractor additionally requires $V_{\rm eff}'(\phi_H)=0$, since $F^2$ vanishes at the horizon, while asymptotic flatness and the massless dilaton assumed in our scattering analysis require $V_{\rm eff}(0)=V_{\rm eff}'(0)=V_{\rm eff}''(0)=0$.

A possible implementation is to combine the one-loop cancellation above with an order-by-order adjustment of the scalar-potential matching conditions in the completed theory. In particular, non-polynomial contributions that cannot be canceled by a finite set of local polynomial counterterms require corresponding cancellations within the additional neutral sector, potentially through further boson--fermion cancellations or a protected flat direction. However, such a construction involves substantial fine-tuning unless the cancellations are symmetry-protected.\footnote{Supersymmetry could enforce the boson--fermion mass and coupling relations needed to maintain these cancellations beyond one loop, reducing the required fine-tuning. If we consider exact unbroken supersymmetry, we would require a supersymmetric completion of both the photon sector and the higher-dimensional EFT operators. In that case, supersymmetric Ward identities would require the all-plus four-photon amplitude to vanish, implying $c^a_{F^4}=c^b_{F^4}$ for the local four-photon operators and hence equal beta functions for their coefficients. Whether a supersymmetric completion of the fermionic operators can realize the negative-running mechanism therefore requires further investigation. If instead we consider that the higher-dimensional EFT operators break supersymmetry, these can generate UV-sensitive local contributions to the scalar potential, so supersymmetry of the scalar--fermion sector alone would not ensure the desired radiative stability.}

In any case, even assuming that the cancellations can be realized so that the EMd backgrounds remain applicable within this restricted class of theories, the conclusion of our analysis does not change. We have shown that the same Yukawa interaction responsible for the negative running also generates a background-dependent fermion mass that obstructs its naive extrapolation to the scales relevant for BH extremality. Consequently, we have found that the negative running alone does not establish a violation of the WGC.

\section{Other implications of a positive four-photon anomalous dimension}
\label{sec:avenues}

Beyond its application to the WGC, the mechanism of negative running compatible with causality and unitarity that we described in Sec.~\ref{sec:F4} is of interest in other areas. Below, we discuss two broader settings in which a positive $\gFa$ and the structure of the corresponding bounds are relevant.

\subsection{Loop-level positivity in the SMEFT}

Positivity bounds have been applied extensively to dimension-eight operators in the SMEFT, in particular to quartic gauge couplings~\cite{Remmen1,Zhang1,Bi1,Yamashita1}, and they have been extended beyond tree level~\cite{Chala1,Li2}. The one-loop renormalization of dimension-eight operators in the SMEFT is known in detail~\cite{Chala3}, and positivity has been found to restrict the signs of some entries of the corresponding anomalous-dimension matrix. However, a coefficient subject to a positivity bound can still receive renormalization contributions of either sign from coefficients that are themselves not constrained by positivity~\cite{Chala2}. Our result is an explicit realization of this possibility, with an important implication for the interpretation of loop-level positivity bounds: a dimension-eight coefficient that is measured or fitted to be negative at low energies need not signal a violation of causality or unitarity in the UV. It can be the IR end of sign-indefinite running driven by light states, as long as the same states contribute IR terms to the dispersion relations. Positivity must then be applied to the arcs, with the computable IR contributions included~\cite{Bellazzini5}.

\subsection{Light-by-light scattering, light scalars and a hidden scalar}

Light-by-light scattering has been observed in ultraperipheral heavy-ion collisions at the Large Hadron Collider (LHC)~\cite{ATLAS1,CMS1,ATLAS2}. These measurements constrain axion-like particles (ALPs)~\cite{Knapen1,CMS1}, Born--Infeld-type extensions of electromagnetism~\cite{Ellis1}, and the Euler--Heisenberg coefficients, which are also probed at low energies through vacuum magnetic birefringence~\cite{Ejlli1}. Collider tests of positivity have been proposed in several channels~\cite{Bellazzini1,Bellazzini2,Bi1,Yamashita1,Gu1,Li1,Liu1,Fernandez1}. A $CP$-even scalar, such as a dilaton, radion or scalar ALP, realizes the mechanism of Sec.~\ref{sec:F4} when coupled to photons and light fermions as in our theory. It is then natural to ask whether a negative $\cFa$, which would be interpreted as a violation of the assumptions behind positivity, could be measured while the scalar itself remains undetected. If so, negativity would serve as an indirect signature of the light state.

In our setup, however, a simple test shows that the scalar cannot remain hidden. The scalar contributes at tree level to the same four-photon amplitudes as $\cFa$:
\begin{subequations}
    \begin{align}
        \mathcal{A}_4(1_{\gamma^+},2_{\gamma^+},3_{\gamma^+},4_{\gamma^+})&\sim\left(\sq{12}^2\sq{34}^2+\sq{13}^2\sq{24}^2+\sq{14}^2\sq{23}^2\right)\left[8\cFa(S)-8\cFb(S)+\dots\right],\label{eq:Aallplus}\\
        \mathcal{A}_4(1_{\gamma^+},2_{\gamma^-},3_{\gamma^-},4_{\gamma^+})&\sim\sq{14}^2\ang{23}^2\left[-\frac{\kappa^2}{u}+8\cFa(S)+8\cFb(S)+\dots\right],\label{eq:Aelastic}
    \end{align}
\end{subequations}
where subleading loop effects are not shown. Measurements of both processes determine $\cFa$ and $\cFb$ separately. For the scalar to remain hidden behind a negative $\cFa$ at the detection energy $\bar S<\Lambda^2$, a necessary condition is $|8\cFa(\bar S)|>\kappa^2/\bar S$. Defining
\begin{equation}
    \zeta(S)\equiv-S\,\frac{8\cFa(S)}{\kappa^2}\,,
    \label{eq:zeta}
\end{equation}
the scalar can be hidden only if $\zeta(\bar S)>1$. The one-loop bound of Eq.~\eqref{eq:F4a}, however, implies
\begin{equation}
    \zeta(S)<\frac{y^2}{8\pi^2}+\frac{\kappa^2S}{64\pi^2}\,,
    \label{eq:zetabound}
\end{equation}
where the second contribution is small when $\kappa^2S\ll1$. Causality and unitarity thus prevent the scalar from remaining unobserved while a negative $\cFa$ is found; for this to happen, the loop-suppressed negativity must compete with a tree-level effect in the same channel, i.e., we would need to go beyond the perturbative regime of $y\ll1$ and $\kappa^2\bar S\ll1$.

We remark that this argument should be read as a thought experiment. More realistically, the $-\kappa^2/u$ pole has a distinctive angular dependence that a dedicated analysis could identify even when it is subdominant. Moreover, a massless dilaton with $\kappa\gg\MPl^{-1}$ mediates a long-range force much stronger than gravity; such a force is severely constrained by tests of the equivalence principle.

\section{Conclusions}
\label{sec:conclusions}

We have studied one-loop positivity bounds in a theory with a massless scalar coupled to photons through a dilaton-like coupling and to a light neutral fermion via a Yukawa coupling. The central result is that the four-photon coefficient $\cFa$ has a sign-indefinite anomalous dimension, driven by the interference between Yukawa-induced scalar exchange and the operator $(\bar\Psi\Psi)F_{\mu\nu}F^{\mu\nu}$. It can therefore run to negative values in the IR through a non-gravitational effect. This is consistent with causality and unitarity because the fermion bubble in scalar exchange relaxes the bound by a term proportional to $1/S$. We have also obtained the corresponding bounds for the four-scalar and two-scalar--two-photon coefficients, including an inelastic bound.

We have also found that in EMd BHs in the magnetic limit, the substitution of vacuum RG-evolved WCs can make the combination of four-derivative coefficients relevant for the extremality condition negative, suggesting a shift of the charge-to-mass ratio towards the subextremal region. However, the local radial weight of the relevant operators, computed on the classical background, is concentrated in regions where the dilaton-induced fermion mass exceeds the vacuum RG scale at which the combination of WCs turns negative (for WCs of size $\mathcal{O}(\Lambda^{-4})$ at the matching scale). Thus, the apparent challenge to the WGC is obstructed by the BH background itself, through the very same Yukawa interaction responsible for the negative running. Given the suppression of the electromagnetic source in the relevant region, we have tested whether the Coleman--Weinberg potential could be important, finding that its one-loop contribution would dominate the electromagnetic source in the dilaton equation unless cancellations occur. This does not undermine the physical significance of our previous analysis, but restricts its applicability to theories in which the radiatively induced scalar potential is sufficiently suppressed by fine-tuned cancellations to preserve the EMd background.

More broadly, the mechanism analyzed here shows that negative values of dimension-eight coefficients at low energies can be the IR end of sign-indefinite running rather than a sign of a pathological UV completion. This is relevant for loop-level positivity in the SMEFT and for light-by-light scattering in the presence of light scalars. In the latter case, we have shown that a light dilaton cannot hide behind an apparently negative four-photon coefficient, since the loop-suppressed negativity must compete with tree-level dilaton exchange. It would be interesting to identify realizations of this mechanism in realistic extensions of the Standard Model.

\begin{acknowledgments}
    I am deeply grateful to Javi Serra for his constant support, encouragement and many insightful discussions. I also thank Brando Bellazzini and the Institut de Physique Théorique (IPhT) in Saclay, where part of this research was carried out, for their hospitality, and Eduardo Velasco-Aja for useful conversations. J.F. acknowledges financial support from the predoctoral contract PRE2021-099269, associated with project CEX2020-001007-S-21-4 under the Severo Ochoa programme, and from grant PID2022-142545NB-C22, both funded by MICIU/AEI/10.13039/501100011033 and co-funded by ESF+ and ERDF/EU, respectively.
\end{acknowledgments}

\appendix

\section{The EFT up to dimension eight}
\label{app:eft}

We characterize the theory of photons $\gamma$, a massless scalar $\phi$ and two massless Weyl fermions $\psi,\chi$, combined into a Dirac spinor $\Psi\equiv(\psi_\alpha,\tilde\chi^{\dot\alpha})$, in terms of the independent on-shell amplitudes generated by operators of dimension eight or less. None of the particles is electrically charged. We take $\MPl\gg\Lambda$ and omit gravity, and we impose $CP$. Independent amplitudes are constrained by little-group scaling, dimensional analysis, Bose and Fermi statistics, and momentum conservation. We use the all-incoming convention for external momenta and spinor-helicity variables for the polarizations.

Since dimension-eight operators containing fermions do not enter the extremality condition, and amplitudes generated by more than two insertions of the Yukawa (for example, from operators like $(\bar\Psi\Psi)^2$) give subleading contributions to our one-loop results, we neglect them. The only on-shell 3-point amplitudes and their corresponding operators are
\begin{subequations}
    \begin{alignat}{1}
        \mathcal{A}_3(1_\phi,2_{\psi^+},3_{\chi^+}) = y\,\sq{23}\,;
        &\qquad y\,\phi\,(\bar\Psi\Psi)\subset\mathcal{L}\,,
        \label{eq:A3yuk}\\
        \mathcal{A}_3(1_\phi,2_{\gamma^+},3_{\gamma^+}) = -\kappa\,\sq{23}^2\,;
        &\qquad -\frac14e^{-2\kappa\phi}F_{\mu\nu}F^{\mu\nu}\subset\mathcal{L}\,.
        \label{eq:A3dil}
    \end{alignat}
\end{subequations}
These are the minimal interactions of Eq.~\eqref{eq:Smin}. The EFT effects arise in 4-point amplitudes. The relevant independent contact amplitudes up to dimension eight and their corresponding operators are
\begin{subequations}
    \begin{alignat}{1}
        \mathcal{A}_4(1_\phi,2_\phi,3_\phi,4_\phi)_c
        = 2\cD(s^2+t^2+u^2)\,;
        &\qquad \cD\,e^{-2\kappa\phi}(\partial_\mu\phi\partial^\mu\phi)^2\subset\mathcal{L}\,,
        \label{eq:A4phi4}\\
        \mathcal{A}_4(1_\phi,2_\phi,3_{\psi^+},4_{\chi^+})_c
        = -\cDP\,\sq{34}\,s\,;
        &\qquad \cDP\,e^{-\kappa\phi}(\partial_\mu\phi\partial^\mu\phi)(\bar\Psi\Psi)\subset\mathcal{L}\,,
        \label{eq:A4phiphiPsiPsi}\\
        \mathcal{A}_4(1_\phi,2_\phi,3_{\gamma^+},4_{\gamma^+})_c
        \supset 2\cDFa\,\sq{34}^2s\,;
        &\qquad \cDFa\,e^{-4\kappa\phi}\partial_\mu\phi\partial^\mu\phi F_{\nu\rho}F^{\nu\rho}\subset\mathcal{L}\,,
        \label{eq:A4phiphigg_a}\\
        \mathcal{A}_4(1_\phi,2_\phi,3_{\gamma^+},4_{\gamma^+})_c
        \supset \frac12\cDFb\,\sq{34}^2s\,;
        &\qquad \cDFb\,e^{-4\kappa\phi}\partial_\mu\phi\partial_\nu\phi F^{\mu\rho}F^\nu_{\;\;\rho}\subset\mathcal{L}\,,
        \label{eq:A4phiphigg_b}\\
        \mathcal{A}_4(1_\phi,2_\phi,3_{\gamma^+},4_{\gamma^-})_c
        = \cDFb\,\sq{23}^2\ang{24}^2\,;
        &\qquad \cDFb\,e^{-4\kappa\phi}\partial_\mu\phi\partial_\nu\phi F^{\mu\rho}F^\nu_{\;\;\rho}\subset\mathcal{L}\,,
        \label{eq:A4phiphigg_opposite}\\
        \mathcal{A}_4(1_{\psi^+},2_{\chi^+},3_{\psi^-},4_{\chi^-})_c
        \supset -4c_{\Psi^4}\sq{12}\ang{34}\,;
        &\qquad c_{\Psi^4}(\bar\Psi\gamma_\mu\Psi)(\bar\Psi\gamma^\mu\Psi)\subset\mathcal{L}\,,
        \label{eq:A4Psi4}\\
        \mathcal{A}_4(1_{\psi^+},2_{\chi^+},3_{\gamma^+},4_{\gamma^+})_c
        = -2\cPF\,\sq{12}\sq{34}^2\,;
        &\qquad \cPF\,e^{-3\kappa\phi}(\bar\Psi\Psi)F_{\mu\nu}F^{\mu\nu}\subset\mathcal{L}\,,
        \label{eq:A4PsiF}\\
        \mathcal{A}_4(1_{\gamma^+},2_{\gamma^+},3_{\gamma^+},4_{\gamma^+})_c
        \supset 8\cFa\sq{12}^2\sq{34}^2+\text{cross.}\,;
        &\qquad \cFa\,e^{-6\kappa\phi}(F_{\mu\nu}F^{\mu\nu})^2\subset\mathcal{L}\,,
        \label{eq:A4g4_allplus_a}\\
        \mathcal{A}_4(1_{\gamma^+},2_{\gamma^+},3_{\gamma^+},4_{\gamma^+})_c
        \supset -8\cFb\sq{12}^2\sq{34}^2+\text{cross.}\,;
        &\qquad \cFb\,e^{-6\kappa\phi}(F_{\mu\nu}\tilde F^{\mu\nu})^2\subset\mathcal{L}\,,
        \label{eq:A4g4_allplus_b}\\
        \mathcal{A}_4(1_{\gamma^+},2_{\gamma^+},3_{\gamma^-},4_{\gamma^-})_c
        \supset 8\cFa\sq{12}^2\ang{34}^2\,;
        &\qquad \cFa\,e^{-6\kappa\phi}(F_{\mu\nu}F^{\mu\nu})^2\subset\mathcal{L}\,,
        \label{eq:A4g4_elastic_a}\\
        \mathcal{A}_4(1_{\gamma^+},2_{\gamma^+},3_{\gamma^-},4_{\gamma^-})_c
        \supset 8\cFb\sq{12}^2\ang{34}^2\,;
        &\qquad \cFb\,e^{-6\kappa\phi}(F_{\mu\nu}\tilde F^{\mu\nu})^2\subset\mathcal{L}\, ,
        \label{eq:A4g4_elastic_b}
    \end{alignat}
\end{subequations}
where $s=(p_1+p_2)^2$, $t=(p_1+p_3)^2$ and $u=(p_1+p_4)^2$. The four-fermion amplitude, Eq.~\eqref{eq:A4Psi4},  is not used in this work. Higher-point amplitudes are not relevant here.

\section{Derivation of the positivity bounds}
\label{app:bounds}

\begin{figure}[t]
    \centering
    \includegraphics[width=0.45\textwidth]{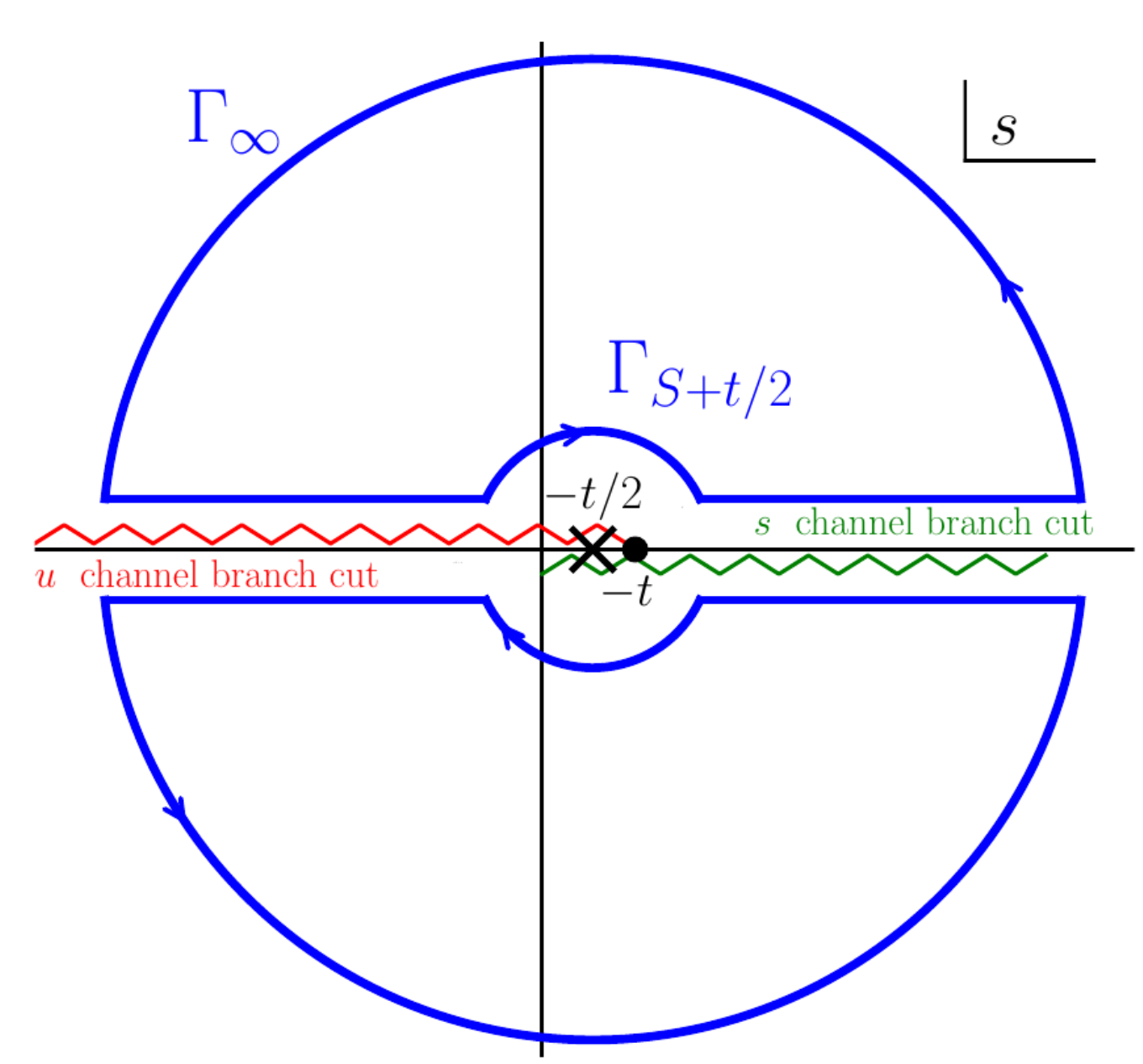}
    \caption{Integration contour and non-analyticities of the amplitudes in the complex-$s$ plane at fixed $t<0$. The contour consists of two arcs of infinite radius, $\Gamma_\infty$; two arcs of radius $S+t/2$ centred at $s=-t/2$, $\Gamma_{S+t/2}$; and paths along the discontinuities from $s\to-\infty$ to $s=-S-t$ and from $s=S$ to $s\to\infty$. The discontinuities are generated at one loop by the $s$- and $u$-channel branch cuts. The subtraction point is at $s=-t/2$.}
    \label{fig:contour}
\end{figure}

Consider an amplitude $\mathcal{A}_4(s,t)$ and the contour $\Gamma$ in the complex $s$ plane of Fig.~\ref{fig:contour}. Then
\begin{equation}
    \frac{1}{2\pi i}\oint_\Gamma\frac{\mathcal{A}_4(s',t)}{(s'+\frac t2)^n}ds'=0
\end{equation}
for some positive integer $n$. We further assume that the amplitudes satisfy the Froissart bound,
\begin{equation}
    \lim_{s\to\infty}\frac{\mathcal{A}_4(s,t)}{s^2}=0\, ,
    \label{eq:Froissart}
\end{equation}
which is strictly true for theories with a mass gap~\cite{Froissart1,Martin2}. In our case, there is no rigorous proof, but there are results suggesting that a polynomial bound of this kind may hold~\cite{Haring1}, as well as explicit examples in perturbative string theory~\cite{Adams1}. Under this assumption, for $n>2$, the integral over the arcs at infinity $\Gamma_\infty$ vanishes and
\begin{equation}
    -\frac{1}{2\pi i}\int_{\Gamma_S}\frac{\mathcal{A}_4(s',t)}{(s'+\frac t2)^n}ds'=\frac1\pi\int_{-\infty}^{-S-t}\frac{{\rm Im}\mathcal{A}_4(s',t)}{(s'+\frac t2)^n}ds'+\frac1\pi\int_S^\infty\frac{{\rm Im}\mathcal{A}_4(s',t)}{(s'+\frac t2)^n}ds'\,.
    \label{eq:dispersion}
\end{equation}
The left-hand side can be computed within the EFT for $S<\Lambda^2$. To obtain Eqs.~\eqref{eq:F4b}, \eqref{eq:F4a}, \eqref{eq:phi4} and \eqref{eq:DFb}, we evaluate it with the one-loop amplitudes of App.~\ref{app:oneloop}. The right-hand side has positivity properties in the forward limit $t\to0$ that follow from the generalized optical theorem. To fully uncover them, we use the partial-wave decomposition
\begin{equation}
    \mathcal{A}_4(1_{h_1},2_{h_2},3_{h_3},4_{h_4})=16\pi\sum_J(2J+1)\,c_J^{h_1h_2h_3h_4}(s)\,d^J_{h_{12},h_{43}}\!\left(1+\frac{2t}{s}\right),
\end{equation}
where $h_{ij}=h_i-h_j$, $c_J$ are the partial waves and $d^J_{m,m'}$ are Wigner $d$-functions. The generalized optical theorem then states that
\begin{equation}
    {\rm Im}\,c_J^{h_1h_2h_3h_4}(s)=\sum_Xa_J^{h_1h_2\to X}(s)\left(a_J^{-h_3,-h_4\to X}(s)\right)^*,
\end{equation}
where the sum runs over all intermediate states $X$ in $1,2\to X\to3,4$ with fixed angular momentum $J$.

\paragraph*{Four scalars.} At tree level, applying the $n=3$ dispersion relation to Eq.~\eqref{eq:IRdiv}, changing variables $s'\to-s'-t$ in the first integral of Eq.~\eqref{eq:dispersion}, and taking $t\to0$ gives
\begin{equation}
    2\cD=16\int_S^\infty\frac{ds'}{(s')^3}\sum_{J,X}(2J+1)\,|a_J^{00\to X}(s')|^2d^J_{0,0}(1)>0\,,
    \label{eq:B6}
\end{equation}
since $d^J_{0,0}(1)=1$. At one loop, the left-hand side contains IR divergences, Eq.~\eqref{eq:C1}, and the naive bound $2\cD[1-\frac{y^2}{4\pi^2}\log(\mu^2/m_{\rm IR}^2)]+\dots>0$ is meaningless. However, in light of the evidence in~\cite{Gervais1,Baratella2}, we can motivate the assumption that collinear divergences due to external-leg corrections factorize on both sides of the dispersion relation,
\begin{equation}
    2\cD(S)F^4(m_{\rm IR})=16\int_S^\infty\frac{ds'}{(s')^3}\sum_{J,X}(2J+1)F^4(m_{\rm IR})\,|\tilde a_J^{00\to X}(s')|^2d^J_{0,0}(1)\,,
\end{equation}
where $F(m_{\rm IR})$ is independent of the kinematics, with one power for each external scalar. The one-loop correction to the bound then amounts to considering the running of $\cD(S)$, Eq.~\eqref{eq:phi4}.

\paragraph*{Four photons.} We apply the $n=1$ dispersion relation to the helicity-stripped amplitudes $\mathcal{A}_4(1_{\gamma^+},2_{\gamma^-},3_{\gamma^-},4_{\gamma^+})/(\sq{14}^2\ang{23}^2)$ and $\mathcal{A}_4(1_{\gamma^+},2_{\gamma^+},3_{\gamma^+},4_{\gamma^+})/(\sq{12}^2\sq{34}^2+\dots)$. The helicity prefactors scale as $s^2$ for $s\to\infty$, so the arcs at infinity vanish. At tree level,
\begin{subequations}
    \label{eq:B9}
    \begin{align}
        8\cFa+8\cFb&=16\int_S^\infty\frac{ds'}{(s')^3}\sum_J(2J+1)\Big[\sum_X|a_J^{+-\to X}(s')|^2+\sum_X|a_J^{++\to X}(s')|^2\Big]\,, \label{eq:B9a}\\
        8\cFa-8\cFb&=16\int_S^\infty\frac{ds'}{(s')^3}\sum_J(2J+1)\Big[\sum_Xa_J^{++\to X}(s')\left(a_J^{--\to X}(s')\right)^*\Big]\,,
    \end{align}
\end{subequations}
where we used $d^J_{0,0}(1)=d^J_{2,2}(1)=1$. For parity eigenstates, $a_J^{--\to X}=a_J^{++\to X}$ if $X$ is even and $a_J^{--\to X}=-a_J^{++\to X}$ if it is odd. Combining the dispersion relations then bounds each coefficient separately, giving Eq.~\eqref{eq:F4treesep}. Loop corrections are included by evaluating the left-hand side with Eqs.~\eqref{eq:C3} and~\eqref{eq:C5}, which contain no IR divergences, giving Eqs.~\eqref{eq:F4b} and~\eqref{eq:F4a}.

\paragraph*{Two scalars and two photons.} The bound on $\cDFb$ follows from the $n=1$ dispersion relation for $\mathcal{A}_4(1_{\gamma^+},2_\phi,3_{\gamma^-},4_\phi)/(\sq{12}^2\ang{23}^2)$. In the forward limit,
\begin{equation}
    \cDFb=-32\int_S^\infty\frac{ds'}{(s')^3}\sum_{J,X}(2J+1)|a_J^{+0\to X}(s')|^2<0\,.
    \label{eq:B10}
\end{equation}
The upper bound is a consequence of the forward limit of the polarizations, $\lim_{t\to0}\sq{12}^2\ang{23}^2\to-s^2$. The one-loop version, Eq.~\eqref{eq:DFb}, uses Eq.~\eqref{eq:C7}, with the IR divergences treated as in the four-scalar case.

For the inelastic bound~\cite{Trott1}, we apply the $n=2$ dispersion relation to $\mathcal{A}_4(1_{\gamma^+},2_{\gamma^+},3_\phi,4_\phi)/\sq{12}^2$ at tree level:
\begin{align}
    2\cDFa+\frac12\cDFb=16\int_S^\infty\frac{ds'}{(s')^3}\sum_J(2J+1)\Big[&\sum_Xa_J^{++\to X}(s')\left(a_J^{00\to X}(s')\right)^*\nonumber\\
    +&\sum_Xa_J^{+0\to X}(s')\left(a_J^{0-\to X}(s')\right)^*\Big]\,.
    \label{eq:B11}
\end{align}
For parity eigenstates, the absolute value of the last term is bounded by $-\cDFb/2$, as follows from Eq.~\eqref{eq:B10}. For the first term, we apply the Cauchy--Schwarz inequality,
\begin{align}
    16\int_S^\infty\frac{ds'}{(s')^3}\sum_J(2J+1)\Big[&\sum_Xa_J^{++\to X}\left(a_J^{00\to X}\right)^*\Big]\equiv\ang{a^{00}_J,a^{++}_J}\,,\nonumber\\
    \left|\ang{a^{00}_J,a^{++}_J}\right|\le&\sqrt{\ang{a^{00}_J,a^{00}_J}\ang{a^{++}_J,a^{++}_J}}\,.
\end{align}
The quantities on the right-hand side are dispersive integrals of elastic amplitudes: $\ang{a^{00}_J,a^{00}_J}$ is given by Eq.~\eqref{eq:B6}, and $\ang{a^{++}_J,a^{++}_J}$ is bounded by Eq.~\eqref{eq:B9a}. Moreover, in our theory, pairs of $\phi$ couple only to parity-even states $X$, so only the even part $\ang{a^{++}_J,a^{++}_J}_{\rm even}$ is needed. $\ang{a^{++}_J,a^{++}_J}_{\rm even}$ is itself bounded by the combination of dispersions that involves only even intermediate states, i.e., by the sum of the two lines of Eq.~\eqref{eq:B9}, which depends only on $\cFa$. Applying Cauchy--Schwarz gives the two-sided constraint of Eq.~\eqref{eq:DFinel0}.

At one loop, we use Eq.~\eqref{eq:C9} to obtain Eq.~\eqref{eq:DFinel1}, and the IR divergences are treated as before. Notice that the two-scalar--two-photon amplitudes contain a factor $F^2(m_{\rm IR})$, precisely as $\sqrt{\ang{a^{00}_J,a^{00}_J}}\sim F^2(m_{\rm IR})$, so the IR collinear factors cancel.

\section{One-loop expressions}
\label{app:oneloop}

Here we collect the one-loop amplitudes used in Sec.~\ref{sec:positivity}. We use 2-, 3- and 4-particle cuts, which give the coefficients of massive bubbles, triangles and boxes in the Passarino--Veltman decomposition. Reviews of these methods can be found in Refs.~\cite{Bern1,Britto1}. Unitarity cuts determine the full one-loop amplitude up to rational terms, which we fix by requiring the cancellation of higher-order poles in momenta~\cite{Chen1}. The collinear IR divergences are obtained by studying their cancellation at the cross-section level against contributions from 5-point amplitudes containing pairs of collinear fermions.

The four-scalar amplitude is
\begin{align}
    \mathcal{A}_4(1_\phi,2_\phi,3_\phi,4_\phi&)_{\rm 1-loop}=\nonumber\\
    &\quad 2\cD(\mu^2)(s^2+t^2+u^2)\left[1-\frac{y^2}{4\pi^2}\log\left(\frac{\mu^2}{m_{\rm IR}^2}\right)\right]\nonumber\\
    &+\frac{c_s^{4\phi}}{2}\log\left(\frac{-s}{\mu^2}\right)+\frac{c_t^{4\phi}}{2}\log\left(\frac{-t}{\mu^2}\right)+\frac{c_u^{4\phi}}{2}\log\left(\frac{-u}{\mu^2}\right)\nonumber\\
    &-\frac{\kappa^4}{16\pi^2}\left[st\left(\log^2\frac st+\pi^2\right)+su\left(\log^2\frac su+\pi^2\right)+tu\left(\log^2\frac tu+\pi^2\right)\right]\nonumber\\
    &+\mathcal{O}\left(\frac{y^4}{16\pi^2}\right),
    \label{eq:C1}
\end{align}
with
\begin{equation}
    c_s^{4\phi}=-\frac{\kappa^4}{8\pi^2}s^2\,,\qquad c_t^{4\phi}=-\frac{\kappa^4}{8\pi^2}t^2\,,\qquad c_u^{4\phi}=-\frac{\kappa^4}{8\pi^2}u^2\,.
\end{equation}
The four-photon elastic amplitude is
\begin{align}
    \mathcal{A}_4(1_{\gamma^+},2_{\gamma^-},3_{\gamma^-},4_{\gamma^+}&)_{\rm 1-loop}=\nonumber\\
    &\quad \sq{14}^2\ang{23}^2\bigg[-\frac{\kappa^2}{u}+8\cFa(\mu^2)+8\cFb(\mu^2)\nonumber\\
    &+\frac{c_s^{\rm el.}}{2}\log\left(\frac{-s}{\mu^2}\right)+\frac{c_t^{\rm el.}}{2}\log\left(\frac{-t}{\mu^2}\right)+\frac{c_u^{\rm el.}}{2}\log\left(\frac{-u}{\mu^2}\right)-\frac{\kappa^2y^2}{8\pi^2u}\log\left(\frac{-u}{\mu^2}\right)\nonumber\\
    &-\frac{\kappa^4}{32\pi^2}\frac{st(s^2+t^2)}{u^4}\left(\pi^2+\log^2\frac st\right)+\frac{\kappa^4}{32\pi^2}\frac{s^2+t^2}{u^2}\bigg]\,,
    \label{eq:C3}
\end{align}
with
\begin{equation}
    c_s^{\rm el.}=\frac{\kappa^4}{48\pi^2}\frac{s(5s^2+4st+11t^2)}{u^3}\,,\quad c_t^{\rm el.}=\frac{\kappa^4}{48\pi^2}\frac{t(11s^2+4st+5t^2)}{u^3}\,,\quad c_u^{\rm el.}=-\frac{\kappa^4}{16\pi^2}+\frac{\cPF\,\kappa y}{\pi^2}\,.
\end{equation}
The four-photon all-plus amplitude is
\begin{align}
    \mathcal{A}_4(1_{\gamma^+},2_{\gamma^+},3_{\gamma^+},4_{\gamma^+})&_{\rm 1-loop}=\nonumber\\
    &\quad \left(\sq{12}^2\sq{34}^2+\sq{13}^2\sq{24}^2+\sq{14}^2\sq{23}^2\right)\left[8\cFa(\mu^2)-8\cFb(\mu^2)\right]\nonumber\\
    &+\sq{12}^2\sq{34}^2\left[-\frac{\kappa^2y^2}{8\pi^2s}+\frac{c_s^{\rm inel.}}{2}\right]\log\left(\frac{-s}{\mu^2}\right)\nonumber\\
    &+\sq{13}^2\sq{24}^2\left[-\frac{\kappa^2y^2}{8\pi^2t}+\frac{c_t^{\rm inel.}}{2}\right]\log\left(\frac{-t}{\mu^2}\right)\nonumber\\
    &+\sq{14}^2\sq{23}^2\left[-\frac{\kappa^2y^2}{8\pi^2u}+\frac{c_u^{\rm inel.}}{2}\right]\log\left(\frac{-u}{\mu^2}\right),
    \label{eq:C5}
\end{align}
with
\begin{equation}
    c_s^{\rm inel.}=c_t^{\rm inel.}=c_u^{\rm inel.}=\frac{\cPF\,\kappa y}{\pi^2}\,.
\end{equation}
The two-scalar--two-photon elastic amplitude is
\begin{align}
    \mathcal{A}_4(1_{\gamma^+},2_\phi,3_{\gamma^-},4_\phi&)_{\rm 1-loop}=\nonumber\\
    &\quad \sq{12}^2\ang{23}^2\bigg[\left(-\kappa^2\frac{t}{su}+\cDFb(\mu^2)\right)\left(1-\frac{y^2}{8\pi^2}\log\left(\frac{\mu^2}{m_{\rm IR}^2}\right)\right)\nonumber\\
    &+\frac{c_s^b}{2}\log\left(\frac{-s}{\mu^2}\right)+\frac{c_t^b}{2}\log\left(\frac{-t}{\mu^2}\right)+\frac{c_u^b}{2}\log\left(\frac{-u}{\mu^2}\right)\nonumber\\
    &+\frac{\kappa^4}{32\pi^2}\frac{st}{u^2}\left(\pi^2+\log^2\frac st\right)+\frac{\kappa^4}{32\pi^2}\frac{tu}{s^2}\left(\pi^2+\log^2\frac tu\right)\bigg]\,,
    \label{eq:C7}
\end{align}
with
\begin{equation}
    c_s^b=\frac{\kappa^4}{24\pi^2}\frac{3t+u}{u}\,,\qquad c_t^b=\frac{\kappa^4}{8\pi^2}\frac{t^2}{su}\,,\qquad c_u^b=\frac{\kappa^4}{24\pi^2}\frac{3t+s}{s}\,.
\end{equation}
Finally, the two-scalar--two-photon amplitude with photons of the same helicity is
\begin{align}
    \mathcal{A}_4(1_{\gamma^+},2_{\gamma^+},3_\phi,4_\phi&)_{\rm 1-loop}=\nonumber\\
    &\quad \sq{12}^2\bigg[s\left(2\cDFa(\mu^2)+\frac12\cDFb(\mu^2)\right)\left(1-\frac{y^2}{8\pi^2}\log\frac{\mu^2}{m_{\rm IR}^2}\right)\nonumber\\
    &+\frac{c_s^{a+b}}{2}\log\left(\frac{-s}{\mu^2}\right)+\frac{c_t^{a+b}}{2}\log\left(\frac{-t}{\mu^2}\right)+\frac{c_u^{a+b}}{2}\log\left(\frac{-u}{\mu^2}\right)\bigg]\,,
    \label{eq:C9}
\end{align}
with
\begin{equation}
    c_s^{a+b}=\frac{\cDP\,\kappa y}{4\pi^2}s\,,\qquad c_t^{a+b}=0\,,\qquad c_u^{a+b}=0\,.
\end{equation}

\section{Black-hole profiles and radial contributions}
\label{app:profiles}

We use the conventions of Eq.~\eqref{eq:Smin}, with $\phi_\infty=0$ and $\MPl=1$. In the following, $R$ denotes the areal radius, so that the symmetry spheres have area $4\pi R^2$. For a regular extremal dyon, we use the horizon radius and the attractor value to define the following parameters~\cite{Loges1,Cremonini1}
\begin{equation}
    R_H=L=\sqrt{|q_eq_m|}\,,\qquad
    e^{-2\kappa\phi_H}=\frac{|q_e|}{|q_m|}=\frac1\xi\,.
    \label{eq:appendixattractor}
\end{equation}
Regarding the electromagnetic field, Maxwell's equations fix $F_{\theta\varphi}=q_m\sin\theta$ and $e^{-2\kappa\phi}R^2F^{tr}=q_e$ in a radial coordinate satisfying $g_{tt}g_{rr}=-1$. Hence,
\begin{equation}
    F^2=\frac{2}{R^4}\left(q_m^2-q_e^2e^{4\kappa\phi}\right)\,,\qquad
    \left.F^2\right|_{R=L}=0\,,
    \label{eq:appendixdyonicF2}
\end{equation}
which we use in Sec.~\ref{sec:bhbackground}. The second equality holds for the regular dyonic horizon, by Eq.~\eqref{eq:appendixattractor}.

\paragraph*{Purely magnetic solution.}
For $q_e=0$, the extremal solution can be written~\cite{Gibbons1,Garfinkle1,Loges1}, in our mostly-minus conventions, as
\begin{equation}
    \begin{aligned}
        ds^2&=\left(1-\frac{r_0}{r}\right)^{\frac{2}{1+2\kappa^2}}dt^2
        -\left(1-\frac{r_0}{r}\right)^{-\frac{2}{1+2\kappa^2}}dr^2
        -R(r)^2d\Omega^2\,,\\
        R(r)^2&=r^2\left(1-\frac{r_0}{r}\right)^{\frac{4\kappa^2}{1+2\kappa^2}}\,,\\
        e^{-2\kappa\phi(r)}&=\left(1-\frac{r_0}{r}\right)^{\frac{4\kappa^2}{1+2\kappa^2}}\,,
    \end{aligned}
    \label{eq:magneticprofile}
\end{equation}
where
\begin{equation}
    r_0=|q_m|\sqrt{\frac{1+2\kappa^2}{2}}\,.
\end{equation}
Here, $r$ is the radial coordinate and the horizon is at $r=r_0$, but $R(r_0)=0$; purely magnetic BHs feature vanishing horizon area.

The two-derivative extremal mass is
\begin{equation}
    M_0=\frac{4\pi q_m^2}{r_0}\,.
    \label{eq:extmagneticmass}
\end{equation}

The invariants needed to evaluate the operators in Eq.~\eqref{eq:LEFT} are
\begin{equation}
    \begin{aligned}
        \phi(r)&=-\frac{2\kappa}{1+2\kappa^2}\log\!\left(1-\frac{r_0}{r}\right),\\
        F^2&=\frac{2q_m^2}{R^4}\,,\\
        (\partial\phi)^2&=-\left(\frac{2\kappa r_0}{1+2\kappa^2}\right)^2
        \frac{e^{2\kappa\phi}}{r^4},\\
        \sqrt{-g}&=R^2\sin\theta\,.
    \end{aligned}
    \label{eq:magneticinvariants}
\end{equation}

\paragraph*{Radial integrals and magnetic extremality.}
Following the method of~\cite{Loges1}, the leading higher-derivative correction to the extremality condition can be obtained by evaluating the corresponding terms of the Euclidean action on the uncorrected non-extremal background, using the thermodynamic relations and subsequently taking the extremal limit at fixed charge. For the purely magnetic solution, this leads to the radial integrals of the four-derivative operators given below.

More concretely, substituting Eq.~\eqref{eq:magneticinvariants} into the three relevant operators of Eq.~\eqref{eq:LEFT} and integrating over the angles, we obtain
\begin{subequations}
    \label{eq:magneticspatial}
    \begin{align}
        4\pi R^2e^{-6\kappa\phi}(F^2)^2
        &=\frac{16\pi q_m^4}{r^6}\,,\\
        4\pi R^2e^{-4\kappa\phi}(\partial\phi)^2F^2
        &=-\frac{8\pi q_m^2}{r^6}
        \left(\frac{2\kappa r_0}{1+2\kappa^2}\right)^2\,,\\
        4\pi R^2e^{-2\kappa\phi}\big[(\partial\phi)^2\big]^2
        &=\frac{4\pi}{r^6}
        \left(\frac{2\kappa r_0}{1+2\kappa^2}\right)^4\,.
    \end{align}
\end{subequations}
Thus, we find the same radial dependence $dr/r^6$ for the three operators. Using Eq.~\eqref{eq:extmagneticmass}, we recover the magnetic extremality correction~\cite{Loges1}:
\begin{align}
    \frac{4\pi|q_m|}{M}\sqrt{\frac{2}{1+2\kappa^2}}\,\bigg|_{\rm ext}-1
    =&\nonumber\\
    \frac{16}{q_m^2(1+2\kappa^2)^2}&
    \int_{r_0}^{\infty}\!dr\,\frac{r_0^5}{r^6}
    \left[\cFa-\frac{\kappa^2}{1+2\kappa^2}\cDFa
    +\frac{\kappa^4}{(1+2\kappa^2)^2}\cD\right]+\dots\,.
    \label{eq:magneticradialexact}
\end{align}
For $\kappa\gg1$, this reduces to Eq.~\eqref{eq:magneticradial}. The identity $\int_{r_0}^{\infty}dr\,r_0^5/r^6=1/5$ then reproduces Eq.~\eqref{eq:ext_mag}.

\paragraph*{Normalized distribution in the areal radius.}
The common normalized weight in the purely magnetic case is
\begin{equation}
    \mathcal P(R)\,d\log(R/L)=5r_0^5\frac{dr}{r^6}\,,\qquad
    \int_0^\infty\!\mathcal P(R)\,d\log(R/L)=1\,.
    \label{eq:magneticweightapp}
\end{equation}
Here $0<R<\infty$ is the full magnetic exterior, and $L$ is only the reference length used to compare this solution with a regular dyon. The change of variables follows from Eq.~\eqref{eq:magneticprofile}, with
\begin{equation}
    \frac{d\log R}{dr}=\frac1r+
    \frac{2\kappa^2}{1+2\kappa^2}\frac{r_0}{r(r-r_0)}\,.
\end{equation}
Consequently,
\begin{equation}
    \mathcal P(R)=\left.\frac{5r_0^5}{r^5}
    \left(1+\frac{2\kappa^2}{1+2\kappa^2}\frac{r_0}{r-r_0}\right)^{-1}
    \right|_{r=r(R)}\,.
    \label{eq:magneticweightfull}
\end{equation}
In the large-$\kappa$ limit, $R\simeq r-r_0$ and Eq.~\eqref{eq:magneticweightfull} becomes Eq.~\eqref{eq:magneticweight}. The accumulated weight is
\begin{equation}
    \int_0^R\!\mathcal P(R')\,d\log(R'/L)
    =1-\left(\frac{r_0}{r(R)}\right)^5\,,
    \label{eq:magneticcumulative}
\end{equation}
and it reaches $50\%$ at $r(R_{50})=2^{1/5}r_0$. Finally, the fermion mass in this solution is
\begin{equation}
    m(R)=|y\phi(R)|=
    \frac{2\kappa|y|}{1+2\kappa^2}
    \log\!\left(\frac{r(R)}{r(R)-r_0}\right)
    \simeq\frac{|y|}{\kappa}\log\!\left(1+\frac{r_0}{R}\right)
    \qquad(\kappa\gg1)\,,
    \label{eq:magneticmassprofile}
\end{equation}
which yields Eq.~\eqref{eq:magneticmassmedian} when evaluated at $r=2^{1/5}r_0$.

\paragraph*{Dyonic backgrounds.}
To complete our argument, we need to extend these results to the magnetic limit of the dyonic regular solution and show that the purely magnetic solution is a good approximation. For a generic nonzero $q_e$, however, the extremal dyonic solution does not admit a simple closed-form profile for arbitrary $\kappa$~\cite{Loges1,Cremonini1}. Here, we use the radial gauge
\begin{equation}
    ds^2=\frac{\rho^2}{R(\rho)^2}dt^2
    -\frac{R(\rho)^2}{\rho^2}d\rho^2-R(\rho)^2d\Omega^2\,,
    \label{eq:dyonicradialgauge}
\end{equation}
where $\rho=0$ at the extremal horizon (where $R(0)=L$). This coordinate $\rho$ is distinct from the magnetic coordinate $r$ above; all comparisons are therefore expressed in terms of the same areal radius $R$. In the conventions of Eq.~\eqref{eq:Smin}, the classical radial equations include
\begin{equation}
    \frac{d^2R}{d\rho^2}=-\frac{R}{2}
    \left(\frac{d\phi}{d\rho}\right)^2\,,\qquad
    \frac{d}{d\rho}\left(\rho^2\frac{d\phi}{d\rho}\right)
    =\frac{\kappa}{R^2}\left(q_e^2e^{2\kappa\phi}
    -q_m^2e^{-2\kappa\phi}\right),
    \label{eq:dyonicradialequations}
\end{equation}
with the remaining Einstein constraint. The regular solution is selected by Eq.~\eqref{eq:appendixattractor} at $\rho=0$ and $\phi\to0$ at infinity~\cite{Loges1,Cremonini1}.

Now, in this gauge, $\sqrt{-g}=R^2\sin\theta$ and
\begin{equation}
    F^2=\frac{2}{R^4}\left(q_m^2-q_e^2e^{4\kappa\phi}\right)\,,\qquad
    (\partial\phi)^2=-\frac{\rho^2}{R^2}
    \left(\frac{d\phi}{d\rho}\right)^2\,.
\end{equation}
The three angularly integrated local densities are therefore
\begin{subequations}
    \label{eq:dyonicspatial}
    \begin{align}
        4\pi R^2e^{-6\kappa\phi}(F^2)^2
        &=\frac{16\pi e^{-6\kappa\phi}}{R^6}
        \left(q_m^2-q_e^2e^{4\kappa\phi}\right)^2\,,\\
        4\pi R^2e^{-4\kappa\phi}(\partial\phi)^2F^2
        &=-\frac{8\pi\rho^2e^{-4\kappa\phi}}{R^4}
        \left(\frac{d\phi}{d\rho}\right)^2
        \left(q_m^2-q_e^2e^{4\kappa\phi}\right)\,,\\
        4\pi R^2e^{-2\kappa\phi}\big[(\partial\phi)^2\big]^2
        &=\frac{4\pi\rho^4e^{-2\kappa\phi}}{R^2}
        \left(\frac{d\phi}{d\rho}\right)^4\,.
    \end{align}
\end{subequations}
In order to compare with the magnetic distributions, each expression must be integrated with $d\rho$, normalized by its own radial integral, and converted to a distribution per unit $d\log(R/L)$ by multiplying by $d\rho/d\log(R/L)$. Notice that, unlike in the purely magnetic case, the three normalized dyonic distributions are not exactly identical.

Fortunately, as shown in Fig.~\ref{fig:F4weight}, the magnetic distributions are indeed a good approximation. For the benchmark of Fig.~\ref{fig:F4weight}, $\kappa=10^5$, $\xi=10^{13}$, $\kappa\Lambda=0.1$, and $y=0.04$, the half-weight areal radii for the $\cFa$, $\cDFa$, and $\cD$ densities are, respectively,
\begin{equation}
    \frac{R_{50}}L=(4.702232,\,4.702267,\,4.702284)\times10^{10}\,.
    \label{eq:dyonicmedians}
\end{equation}
The purely magnetic case yields $R_{50}/L=4.702255\times10^{10}$. Moreover, the relative deviations of the three dyonic distributions from the magnetic distribution are below $4\times10^{-6}$ in the interval containing the central $98\%$ of each integrated numerical density.

Finally, regarding the difficulties related to the numerical integration, if we linearize the scalar equation in Eq.~\eqref{eq:dyonicradialequations} around the regular attractor, we obtain
\begin{equation}
    \frac{d}{d\rho}\left(\rho^2\frac{d(\phi-\phi_H)}{d\rho}\right)
    =4\kappa^2(\phi-\phi_H)+\cdots\,,
\end{equation}
so the regular perturbation behaves as~\cite{Cremonini1,Poletti1,Galtsov1}
\begin{equation}
    \phi-\phi_H\propto\rho^{\frac{\sqrt{1+16\kappa^2}-1}{2}}
    \qquad(\rho\to0)\,.
    \label{eq:dyonicgamma}
\end{equation}
For $\kappa=10^5$ this exponent is approximately $2\times10^5$, making the direct numerical integration extremely sensitive to tiny departures from the attractor. The numerical profiles used in Fig.~\ref{fig:F4weight} are thus directly resolved only for $R/L\gtrsim2.27\times10^4$.

\end{document}